\documentclass[sigconf]{acmart}
\AtBeginDocument{%
  }

\setcopyright{rightsretained} 
\copyrightyear{2024}
\acmYear{2024}

\acmConference[Research in progress]{Draft}{Dec, 2024}{Singapore}

\makeatletter
\gdef\@copyrightpermission{
  \begin{minipage}{0.3\columnwidth}
  \href{https://creativecommons.org/licenses/by/4.0/}{\includegraphics[width=0.90\textwidth]{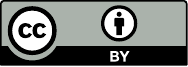}}
  \end{minipage}\hfill
  \begin{minipage}{0.7\columnwidth}
   \href{https://creativecommons.org/licenses/by/4.0/}{This work is licensed under a Creative Commons Attribution International 4.0 License.}
  \end{minipage}
  \vspace{5pt}
}
\makeatother
\usepackage{hyperref}
\usepackage{cleveref}

\usepackage{bbding}

\usepackage{subfigure}

\usepackage{multirow}
\usepackage{threeparttable}

\usepackage{bm}

\usepackage{xspace}
\newcommand{\eg}{\emph{e.g.,}\xspace}
\newcommand{\ie}{\emph{i.e.,}\xspace}

\begin{document}

\title{A Flexible and Scalable Framework for Video Moment Search}

\author{Chongzhi Zhang}
\affiliation{%
  \institution{Nanyang Technological University} 
  \country{Singapore}
  }
  \email{chongzhi001@e.ntu.edu.sg}

\author{Xizhou Zhu}
\affiliation{%
  \institution{SenseTime Research}
  \country{China}}
\email{zhuxizhou@sensetime.com}

\author{Aixin Sun}
\authornote{Corresponding author.}
\affiliation{%
  \institution{Nanyang Technological University}
  \country{Singapore}
  }
\email{axsun@ntu.edu.sg}

\begin{abstract}
Video moment search, the process of finding relevant moments in a video corpus to match a user’s query, is crucial for various applications. Existing solutions, however, often assume a single perfect matching moment, struggle with inefficient inference, and have limitations with hour-long videos. This paper introduces a flexible and scalable framework for retrieving a ranked list of moments from collection of videos in any length to match a text query, a task termed Ranked Video Moment Retrieval (RVMR). Our framework, called Segment-Proposal-Ranking (SPR), simplifies the search process into three independent stages: \textit{segment retrieval}, \textit{proposal generation}, and \textit{moment refinement with re-ranking}. Specifically, videos are divided into equal-length segments with precomputed embeddings indexed offline, allowing efficient retrieval regardless of video length. For scalable online retrieval, both segments and queries are projected into a shared feature space to enable approximate nearest neighbor (ANN) search. Retrieved segments are then merged into coarse-grained moment proposals. Then a refinement and re-ranking module is designed to reorder and adjust timestamps of the coarse-grained proposals. Evaluations on the TVR-Ranking dataset demonstrate that our framework achieves state-of-the-art performance with significant reductions in computational cost and processing time. The flexible design also allows for independent improvements to each stage, making SPR highly adaptable for large-scale applications.\footnote{Codes are available at \href{https://github.com/Ranking-VMR/SPR}{https://github.com/Ranking-VMR/SPR}.}
\end{abstract}


\maketitle

%======================
\section{Introduction}
\label{sec:intro}
%======================

Through text queries, users can access various online resources, such as web pages, images, and videos. However, retrieving relevant moments from a large video corpus remains a challenge. These retrieved moments can be valuable for tasks like video editing, identifying scenes in surveillance footage~\cite{DBLP:conf/cvpr/YuanZLLCJJ24}, and finding segments about specific topics in educational videos~\cite{DBLP:conf/coling/GuptaAD24}, among others.

\begin{figure}[t!]
    \centering
    \includegraphics[trim={2.7cm 8.2cm 13cm 3cm},clip, width=\columnwidth]{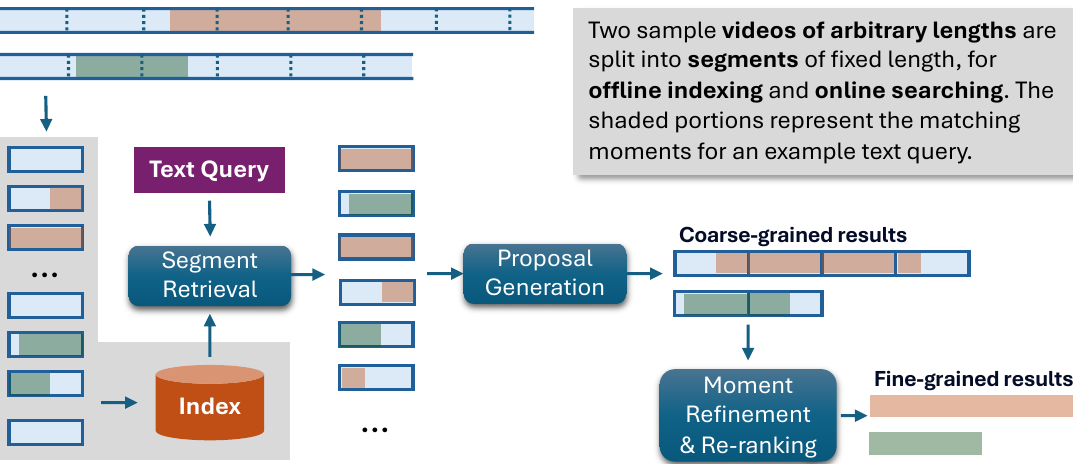}
    \caption{\textbf{The Segment-Proposal-Ranking (SPR) framework.} All videos are divided into non-overlapping, equal-length segments (\eg, 4 seconds) for indexing and searching. The final results are computed based on the relevant segments retrieved.}
    \label{fig:overview}
\end{figure}

Formally, the task of retrieving a ranked list of video moments from a video corpus for a text query is known as Ranked Video Moment Retrieval (RVMR)~\cite{liang2024tvrrankingdatasetrankedvideo}. In the CV and NLP communities, several related tasks have been explored, including Natural Language Video Localization (NLVL)~\cite{DBLP:conf/iccv/GaoSYN17,DBLP:conf/iccv/KrishnaHRFN17}, which involves locating \textit{one specific moment} within one input video based on a text query, and Video Corpus Moment Retrieval (VCMR)~\cite{DBLP:journals/corr/abs-1907-12763}, which focuses on retrieving \textit{one particular moment} from a collection of videos. However, user queries in typical searches may not always provide detailed descriptions about one and only one moment. More importantly, multiple matching moments may exist, each with varying degrees of relevance to the query, similar to traditional web searches. Aiming to retrieve one specific moment, all existing solutions for NLVL and VCMR are designed to take an entire video as input, making it challenging to process hour-long videos. To the best of our knowledge, no \textit{practical solutions} have been proposed for RVMR. By `practical', we mean a solution that: (i) processes a query in a reasonable time, ideally in real-time; (ii) handles videos of any length, from seconds to hours; and (iii) is capable of searching a large collection of videos, in the scale of thousands or even millions. 

In this paper, we propose a simple, flexible, and scalable framework for RVMR, called SPR. As illustrated in~\cref{fig:overview}, all videos are divided into non-overlapping, equal-length segments (\eg 4 seconds), with their embedding features pre-computed and indexed offline. Given a text query, we retrieve relevant segments and merge them into proposals based on their original positions in the source videos. These proposals serve as coarse-grained results. Finally, a refinement and re-ranking process is applied to the proposals to locate matching moments with more precise timestamps and to rank them by their relevance to the query. We name our framework SPR for \textit{segment-proposal-ranking}.

The use of fixed-length segments offers two advantages. First, it enables the framework to handle videos of any length by dividing them into manageable segments. Second, it standardizes the segment retrieval process, allowing the use of effective visual feature extractors on fixed-length segments. To ensure efficiency and scalability, both the text features from user queries and the visual features from segments are projected into a shared feature space, enabling approximate nearest neighbor (ANN) search through the Faiss Index~\cite{DBLP:journals/corr/abs-2401-08281}. Existing solutions from NLVL and/or VCMR can be adapted to refine video moments. Notably, since refinement is applied only to the coarse-grained results, about 100 proposals containing relevant segments, the computational cost remains low.

We evaluate an instance of our SPR framework on the TVR-Ranking dataset~\cite{liang2024tvrrankingdatasetrankedvideo}, the only dataset annotated for RVMR. Experimental results, measured by NDCG, show that SPR achieves impressive retrieval performance with high efficiency, making it suitable for practical applications. With the refinement and re-ranking module, our implementation reaches state-of-the-art results while maintaining efficient processing times. On average, processing a user query across 20K videos takes less than one second. Additionally, our results demonstrate the framework's robustness, even when numerous unrelated videos (\eg videos from other datasets) are added to the corpus. Most importantly, our framework’s design is highly flexible, allowing independent improvements to each component: segment retrieval, proposal generation, and refinement \& re-ranking.
%======================
\section{Related Work}
\label{sec:related}
%======================

%=====================================
\subsection{Localizing Moment in Video(s)}
%=====================================

Several tasks focus on locating moments within videos based on text queries. The most widely studied task is Natural Language Video Localization (NLVL), also known as Video Moment Retrieval (VMR) and Temporal Sentence Grounding in Video (TSGV). This task involves retrieving a temporal segment within an untrimmed video that semantically matches a natural language query. Early solutions~\cite{DBLP:conf/aaai/Xu0PSSS19,DBLP:conf/aaai/XiaoCZJSYX21,DBLP:conf/mm/QuTZ0DZX20,DBLP:conf/aaai/ZhangPFL20,DBLP:conf/cvpr/WangZL0L21,DBLP:conf/acl/ZhangSJZ20,DBLP:conf/acl/ZhangSJZZG21,DBLP:conf/emnlp/CaoCSZZ21} primarily focused on videos of only a few minutes in length~\cite{DBLP:journals/tacl/RegneriRWTSP13,DBLP:conf/iccv/GaoSYN17,DBLP:conf/iccv/KrishnaHRFN17,DBLP:conf/iccv/HendricksWSSDR17}. Recently, with the introduction of the MAD dataset~\cite{DBLP:conf/cvpr/SoldanPAH0GG22}, a large-scale movie dataset, some works~\cite{DBLP:conf/iccv/BarriosSCHG23,DBLP:conf/acl/HouZ0GYCNSD23,DBLP:conf/iccv/PanHGLSPZ23,10.1007/978-3-031-72664-4_20} have extended the NLVL task to hour-long videos. To tackle the challenge of long-video NLVL efficiently, recent approaches~\cite{DBLP:conf/acl/HouZ0GYCNSD23,DBLP:conf/iccv/PanHGLSPZ23,10.1007/978-3-031-72664-4_20} often divide videos into shorter videos, breaking down moment prediction into two steps, retrieving the most relevant short videos and then applying NLVL within them for precise moment prediction. The key assumption behind is that a short video fully contains a matching moment. This reformulation of long-video NLVL aligns with a related task known as Video Corpus Moment Retrieval (VCMR)~\cite{DBLP:journals/corr/abs-1907-12763}, which involves retrieving a moment from a collection of videos.

While existing models for VCMR~\cite{lei2020tvr,DBLP:conf/emnlp/LiCCGYL20,DBLP:conf/mm/HouNC21,DBLP:conf/sigir/0048SJNZZG21} can perform relatively efficient searches from a corpus to locate a desired moment, they still struggle to function as practical moment search engines due to an unrealistic assumption: the existence of one and only one ``perfect match'' moment for each query. To address this, \citet{liang2024tvrrankingdatasetrankedvideo} introduced Ranked Video Moment Retrieval (RVMR), a task more aligned with real-world search scenarios, along with the corresponding dataset, TVR-Ranking. RVMR emphasizes retrieving a ranked list of video moments that best match an text query across a video collection.

In addition to the flexible design of our SPR framework, there are two key distinctions of our solution from the existing solutions. One is that we do not assume each segment fully contains a matching moment. The other is  the adaptation of approximate nearest neighbor (ANN) search from recent advances in neural information retrieval, enhancing the scalability and practicality of video moment search.

%====================================
\subsection{Retrieval Frameworks}
%====================================

A retrieval framework identifies and retrieves relevant information in response to an information need. It has been widely used for searching various resources, including web pages~\cite{DBLP:conf/wsdm/BenderskyCD11,DBLP:journals/ir/MacdonaldSO13}, text documents~\cite{DBLP:journals/ftir/RobertsonZ09,DBLP:conf/emnlp/KarpukhinOMLWEC20}, images~\cite{DBLP:journals/csur/DattaJLW08,radford2021learning}, videos~\cite{DBLP:conf/mir/MithunLMR18,DBLP:journals/ijon/LuoJZCLDL22}, and more. Retrieval also serves as an essential preliminary step for more sophisticated tasks, such as Retrieval-Augmented Generation (RAG) in text~\cite{DBLP:conf/nips/LewisPPPKGKLYR020,DBLP:conf/icml/BorgeaudMHCRM0L22} and multi-modal~\cite{DBLP:conf/iclr/ChenHSC23,DBLP:conf/iccv/ZhangGPCHLYL23} applications.

A typical retrieval process consists of two main phases: (i) encoding objects for retrieval into a specific representation, and (ii) constructing an index to organize these representations for efficient search. Retrieval methods were initially developed for text documents, with early approaches generally using sparse retrieval techniques, exampled by  BM25~\cite{DBLP:conf/sigir/RobertsonW97,DBLP:conf/sigir/LaffertyZ01,DBLP:journals/ftir/RobertsonZ09}. With the rise of deep learning, dense retrieval methods~\cite{DBLP:conf/emnlp/KarpukhinOMLWEC20,DBLP:conf/sigir/KhattabZ20,DBLP:conf/naacl/GaoDC21} have become the new standard. These models are trained to align query and document representations using specific distance metrics. Specifically, queries and documents are encoded as dense embeddings, and an efficient index~\cite{DBLP:conf/stoc/IndykM98,DBLP:journals/pami/JegouDS11,DBLP:journals/pami/MalkovY20} is built offline to enable fast, scalable online search.

Considerable efforts have focused on multi-modal retrieval, where the main challenge is creating unified embeddings across different modalities. Multi-modal pre-trained alignment models are often used to achieve this goal for various modalities, including image-text~\cite{DBLP:conf/icml/JiaYXCPPLSLD21,radford2021learning}, video-text~\cite{DBLP:conf/emnlp/XuG0OAMZF21,DBLP:conf/cvpr/ChengW0CBB23}, and audio-text~\cite{DBLP:conf/icassp/LouXWY22,DBLP:conf/icassp/WuCZHBD23} alignment. For instance, \citet{DBLP:conf/iclr/LiuXL0023} use CLIP to create a universal multi-modal dense retrieval model, UniVL-DR, which supports both efficient image and text retrievals. Our work is similar in the sense to addresses the multi-modal retrieval problem by building a universal space for visual and text data. 

Unlike typical image-text or video-text retrieval tasks, where cross-modal content is well-aligned (\eg, an image or video matches its caption or description), our task focuses on identifying specific moments within videos based on a query. This means that a query does not align with an entire video, but rather with certain moments that are not pre-annotated. Some videos may contain no such moments, while others may contain many. To make the task even more challenging, the matching moments can vary in length. Because existing solutions developed for multi-modal retrieval come with an underlying assumption to retrieve either an image or a video as a whole, they cannot be directly applied to this new task. As a result, we must balance fine-grained video comprehension with retrieve efficiency.

%======================
\section{The SPR Framework}
\label{sec:Model}
%======================
Following \cref{fig:overview}, we detail the three main components of the SPR framework. For each component, we first describe the generic design and then present our instantiation, \ie, a specific implementation of the component.

\begin{figure}
         \includegraphics[width=\columnwidth]{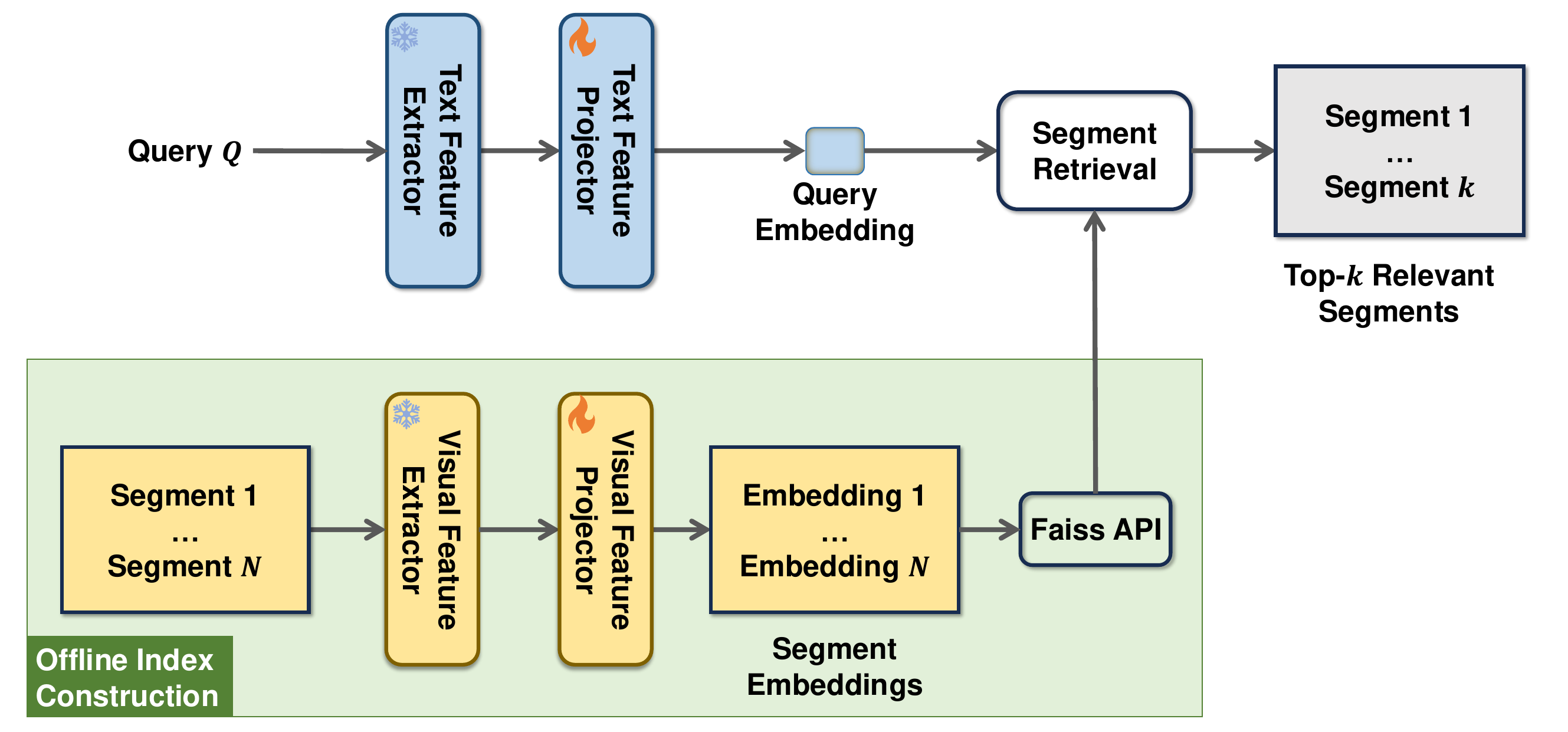}
    \caption{\textbf{Segment retrieval.} With the offline constructed index, the online search/inference takes less than 0.2 seconds to retrieve 100-200 relevant segments for a given query.}

    \label{fig:seg_retrieval}
\end{figure}

%===============================
\subsection{Segment Retrieval} \label{sec:seg_retri}
%===============================

All videos are divided into equal-length segments (\eg, 4 seconds) with precomputed embeddings indexed offline.  During online inference, the top-ranked relevant segments (\eg 100 - 200) are retrieved from the index for a given query. The process is illustrated in \cref{fig:seg_retrieval}.

Formally, given a video corpus, all videos are divided into non-overlapping segments of fixed length $\tau_{S}$. The set of resulting segments is denoted as $\mathcal{S}=\left\{S_{1},S_{2},\dots,S_{|\mathcal{S}|}\right\}$, where $|\mathcal{S}|$ represents the total number of segments. For each segment, its original position (\ie start/end timestamps) and source video are recorded as well.

%==============================
\subsubsection{Offline Index Construction} \label{sec:index_construction}
%==============================

An index is a specialized data structure that stores data embeddings offline, enabling search algorithms to retrieve data instances efficiently. To index a segment, we extract its visual features by using an off the shelf pre-trained model (\eg, CLIP). Similarly, embeddings of the query can be extracted from a pre-trained language model or multi-modality model (\eg BERT or CLIP). Illustrated in \cref{fig:seg_retrieval}, there are two learnable feature projectors (one for the query, and the other for segment),  to project segment and query embeddings to the same feature space. 

Specifically, given a segment, $T$ frames are uniformly sampled and processed by the visual feature extractor to form a feature sequence $\widetilde{\bm{F}^{S}}=\left[\widetilde{\bm{f}_i^{S}}\right]_{i=1}^{T} \in \mathbb{R}^{T \times d_{S}}$. The visual feature projector then applies temporal modeling to the feature sequence, generating a segment embedding $\bm{f}^{S} \in \mathbb{R}^{d}$ that aligns with the query feature space. Note that, these segment embeddings do not interact with the query prior to the search process and are therefore precomputed offline and stored in the index.

After feature projection, both segment and query embeddings reside in the same feature space. We then use cosine similarity as the distance metric between the query and the segment: $d(Q,S)=\frac{{\bm{f}^{Q}}^{\top} \bm{f}^{S}}{\|\bm{f}^{Q}\|\|\bm{f}^{S}\|}$, where $\bm{f}^{Q}$ is the embedding of the query $Q$. Additionally, we can apply nearest neighbor (NN) search, or its more efficient version, approximate nearest neighbor (ANN) search, on these embeddings, provided that we pre-normalize the embeddings of both modalities~\cite{DBLP:conf/icml/NeyshaburS15,DBLP:journals/corr/abs-2207-13443}.

\subsubsection{Online Segment Retrieval} 
Given a query $Q$, the text feature extractor produces either a word embedding sequence $\widetilde{\bm{F}^{Q}}=\left[\widetilde{\bm{f}_i^{Q}}\right]_{i=1}^{N_{Q}} \in \mathbb{R}^{N_{Q} \times d_{Q}}$ or a sentence-level embedding $\widetilde{\bm{f}^{Q}} \in \mathbb{R}^{d_{Q}}$. The text feature projector then generates the query embedding $\bm{f}^{Q}$ aligned with the segment feature space. This embedding is used as input to the search index, which returns the top-$k$ relevant segments. Since dense retrieval has been extensively studied, many off-the-shelf indexes are available, offering high efficiency and scalability. 

%========================================
\subsubsection{Instantiation} \label{sec:inst-seg_retr}
%========================================
In our implementation, for feature extraction across both modalities, we use a frozen CLIP model due to its robust alignment between image and language features. Following \cite{radford2021learning}, we use the output of the $[\texttt{class}]$ token in the visual branch as the frame representation sampled from a video segment. For text, we take activations from the highest Transformer layer at the $[\texttt{EOS}]$ token as the feature representation. This approach encodes the query into a sentence-level feature $\widetilde{\bm{f}^{Q}} \in \mathbb{R}^{d_{Q}}$, while each segment is represented by an image feature sequence of $T$ frames $\widetilde{\bm{F}^{S}}=\left[\widetilde{\bm{f}_i^{S}}\right]_{i=1}^{T} \in \mathbb{R}^{T \times d_{S}}$.

Taking in the image feature sequence, the visual feature projector utilizes a sequential Transformer~\cite{DBLP:journals/ijon/LuoJZCLDL22} with $L_{S}$ layers. A learnable positional embedding layer is added before the Transformer. The frame feature sequence $\widetilde{\bm{F}^{S}}$ is combined with the positional embedding $\bm{P}$ and processed through the Transformer, yielding $\bm{F}^{S}=\operatorname{Transformer}(\widetilde{\bm{F}^{S}}+\bm{P}) \in \mathbb{R}^{T \times d}$. The resulting feature sequence $\bm{F}^{S}$ is aggregated into a segment-level feature using mean pooling: $\bm{f}^{S} = \operatorname{mean-pooling}(\bm{F}_{0}^{S},\dots,\bm{F}_{T-1}^{S}) \in \mathbb{R}^{d}$. The text feature projector is implemented with a simple linear layer, projecting the query feature as $\bm{f}^{Q} = W_{Q} \cdot \widetilde{\bm{f}^{Q}} + b_{Q} \in \mathbb{R}^{d}$. This projected feature $\bm{f}^{Q}$ is then used for matching with segment features. 

For model training, our objective is to improve the both projectors’ capability to distinguish between semantically related query-segment pairs and non-matching pairs. We apply a contrastive learning objective, aligning each query with its matching segments while distancing non-matching pairs. Specifically, for a batch of $B$ query-segment pairs, the model computes $B \times B$ similarity scores. In the RVMR task setting, each row or column may contain one or more positive pairs, with all remaining elements treated as negative samples. Consequently, we employ MIL-NCE~\cite{DBLP:conf/cvpr/MiechASLSZ20} to calculate the contrastive loss in both row-wise and column-wise directions in a batch: 
\begin{align}
    \mathcal{L}_{Q2S} &= -\frac{1}{B} \sum_{i=1}^B \log\left(\frac{ \sum_{j:(Q_i, S_j)\in \mathcal{P}} \exp\left(d(Q_i, S_j)\right)}{\sum_{j=1}^B \exp\left(d(Q_i, S_j)\right)}\right), \notag\\
    \mathcal{L}_{S2Q} &= -\frac{1}{B} \sum_{i=1}^B \log\left(\frac{ \sum_{j:(Q_j, S_i)\in \mathcal{P}} \exp\left(d(Q_j, S_i)\right)}{\sum_{j=1}^B \exp\left(d(Q_j, S_i)\right)}\right), \notag\\
    \mathcal{L} &= \frac{1}{2}\left(\mathcal{L}_{Q2S} + \mathcal{L}_{S2Q}\right),
    \label{eq:loss_segment_retrieval}
\end{align}
where $\mathcal{P}$ denotes the set of positive query-segment pairs.

For dense retrieval index, we use Faiss~\cite{DBLP:journals/corr/abs-2401-08281}, an open-source, production-ready library. 
We experiment with three types of indexes: flat, IVF, and IVFPQ. A flat index directly stores each feature embedding and computes exact distances to all embeddings during search. The IVF and IVFPQ indexes enhance efficiency through Approximate Nearest Neighbor (ANN) search. An IVF index clusters the data points into groups, limiting searches to relevant clusters for faster performance. The IVFPQ index combines clustering with vector compression via product quantization. While IVF and IVFPQ offer increasing search speed, they come with progressively lower accuracy.

\subsection{Coarse Moment Proposal Generation} \label{sec:prop generation}

To generate moment proposals from the top-$k$ most relevant segments, we employ a simple rule-based approach, to merge adjacent segments from the same source video into a single proposal, as shown in \cref{fig:overview}. We have also experimented linking segments separated by a gap smaller than $\tau_{G}$, which brings in negligible improvements. 

The generated proposals can be considered as a set of coarse-grained results to users. The proposal is ranked by the highest rank from among its constituent segments.  However, there are two limitations. First, each proposal consists of one or more segments, with the minimum time scale constrained by the segment length $\tau_{S}$. Second, the rule-based proposal ranking may not align well with the true ranking of the target moments.

\begin{figure}
    \includegraphics[width=\columnwidth]{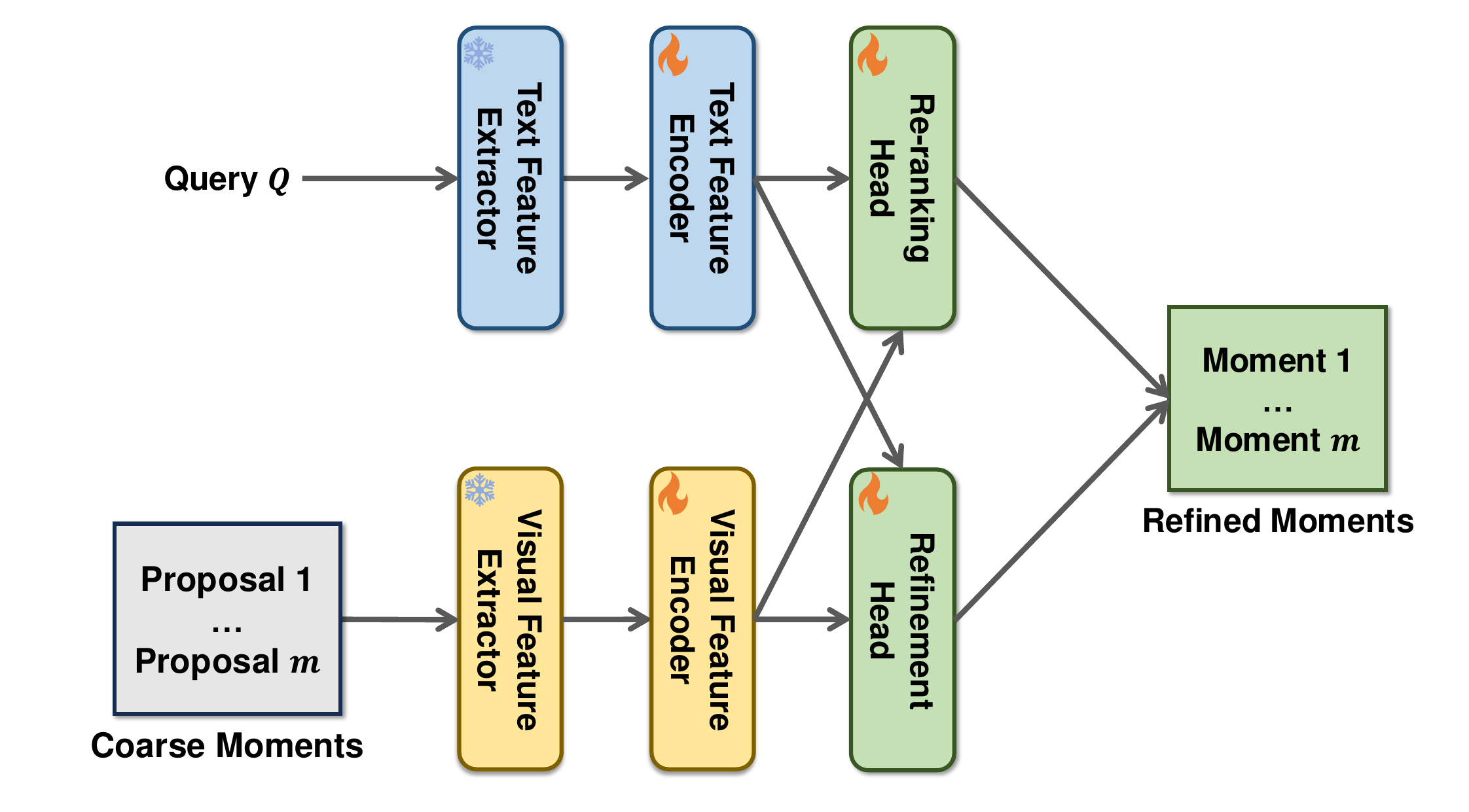}
    \caption{\textbf{Refinement and re-ranking.} This module computes precise timestamps of matching moments and re-ranks them by their relevance to the given query.}
    \label{fig:refinement}
\end{figure}

%===================================
\subsection{Moment Refinement and Re-ranking} \label{sec:ref-rerank-module}
%===================================
To achieve precise moments and ranking as the fine-grained results, a refinement and re-ranking module is applied to the proposals, shown in \cref{fig:refinement}.

%==========================
\subsubsection{Inference Pipeline}
%==========================

Given a coarse proposal $M$, we extract its visual features using a visual extractor, resulting in moment features $\widetilde{\bm{F}^{M}}$. Simultaneously, the query feature $\widetilde{\bm{F}^{Q}}$ or $\widetilde{\bm{f}^{Q}} \in \mathbb{R}^{d_{Q}}$ is generated by a text feature extractor.
\begin{align}
\widetilde{\bm{F}^{M}}&=\left[\widetilde{\bm{f}_i^{M}}\right]_{i=1}^{N_M} \in \mathbb{R}^{N_M \times d_M}, \\ 
\widetilde{\bm{F}^{Q}}&=\left[\widetilde{\bm{f}_i^{Q}}\right]_{i=1}^{N_{Q}} \in \mathbb{R}^{N_{Q} \times d_{Q}}.
\end{align}

The two feature extractors can be the same as those used in \cref{sec:seg_retri} or different ones. Subsequently, a pair of text encoder and visual encoder performs fine-grained modeling on each modality. These dual-modal features are then passed into two heads for \textit{re-ranking} and \textit{moment refinement}, respectively. The re-ranking head computes a query-moment matching score $\varphi(Q,M)$, while the refinement head generates refined start and end boundaries $(t_{s}, t_{e})$ for each moment.

Technically, this module is similar to solutions for NLVL or VCMR. However, the key focus here is to refine the coarse proposals, which already provide good estimates of the final moments. Additionally, proposals are typically much shorter than full videos, leading to more efficient processing. Furthermore, the total number of proposals is relatively small, about 100 in our experiments, because only the top-$k$ more relevant segments are retrieved. All these factors contribute to the efficiency and scalability.

%=======================================
\subsubsection{Instantiation} \label{sec:inst-ref_rerank}
%=======================================

We evaluated two refinement and re-ranking implementations, with main differences in the visual and text feature extractors and projectors.

The first implementation utilizes CLIP feature extractors as well as the text and video feature projectors as described in \cref{sec:seg_retri}. It computes the moment feature sequence $\bm{F}^{M} \in \mathbb{R}^{N_M \times d}$ and query feature $\bm{f}^{Q} \in \mathbb{R}^d$. These features are inputs to both the re-ranking and refinement heads. The re-ranking head computes cosine similarities between $\bm{f}^{Q}$ and $\bm{F}^{M}$, selecting the maximum score in this sequence as the query-moment matching score $\varphi(Q,M)$. The refinement head applies a feed-forward layer and a 1D convolution to generate the refined timestamps $(t_{s}, t_{e})$.

The second implementation is based on ReLoCLNet~\cite{DBLP:conf/sigir/0048SJNZZG21}, following the configurations from~\cite{liang2024tvrrankingdatasetrankedvideo} with features from~\cite{lei2020tvr}. It  utilizes I3D~\cite{carreira2017quo} for visual feature extraction and BERT~\cite{devlin2018bert} for text feature extraction. Additionally, RoBERTa~\cite{liu2019roberta} is used to extract features from video subtitles. Both the video and subtitle features are simultaneously input into the visual encoder for further sequence modeling. As the implementation largely follows ReLoCLNet~\cite{DBLP:conf/sigir/0048SJNZZG21}, we do not detail the model here.

For module training in all settings, we follow the methodology of~\cite{DBLP:conf/sigir/0048SJNZZG21}, applying video contrastive learning and hinge loss on the predicted score $\varphi(Q,M)$, as well as frame contrastive learning and moment localization loss on the predicted start and end boundaries $(t_{s}, t_{e})$. For further details, please refer to~\cite{DBLP:conf/sigir/0048SJNZZG21}. Note that before refinement and re-ranking, we add a contextual padding of length $\tau_{C}$ to all proposals. This padding increases the likelihood of fully capturing relevant moments within the proposals, providing a higher upper bound for refinement.

%======================
\section{Experiments}
\label{sec:exp}

We conduct experiments on \textbf{TVR-Ranking}~\cite{liang2024tvrrankingdatasetrankedvideo}, the only dataset designed for the RVMR task. Next, we detail the dataset, experiment settings, and report results. 

%======================
\subsection{Datasets and Evaluation Metrics}
%======================
TVR-Ranking is developed using the raw videos and existing moment annotations from the TVR dataset~\cite{lei2020tvr}. It draws from 19,614 videos in TVR as its video corpus, rewrites moment descriptions as queries, and provides relevance annotations for matching moments to selected queries. The relevance levels range from 1 (least relevant) to 4 (perfect match). A total of 3,281 queries are manually annotated with relevant moments, resulting in 94,442 query-moment pairs. The annotated queries are divided into 500 validation and 2,781 test queries. The remaining 69,317 queries form a pseudo-training set, created by assigning the top $N$ moments to each query based on query-caption similarity generated by SimCSE~\cite{DBLP:conf/emnlp/GaoYC21}, with $N=40$ in our experiments. The caption here refers to the rewritten moment descriptions with specific details removed \eg, character names. The average video duration is 76.2 seconds, while the average moment duration is 8.7 seconds.

The dataset also comes with an evaluation metric for the RVMR task, $\bm{\textbf{NDCG}@K,\textbf{IoU}\geq \mu}$~\cite{liang2024tvrrankingdatasetrankedvideo}. Normalized Discounted Cumulative Gain (NDCG) is a metric designed to assess ranking results across varying relevance levels. The Discounted Cumulative Gain (DCG) for the top $K$ ranked results is defined as $\text{DCG}@K=\sum_{k=1}^{K}\frac{2^{rel_{i}}-1}{\operatorname{log_2}(i+1)}$, where $rel_{i}$ denotes the relevance level.\footnote{This formulation of DCG, proposed by~\cite{10.1145/1102351.1102363}, places greater emphasis on highly relevant ground truth instances. It is widely applied in industrial contexts, e.g., web search.} Then, $\text{DCG}@K$ is normalized against the DCG of an ideal top $K$ ranking (IDCG), resulting in $\text{NDCG}@K$. For a given $\textbf{IoU}$ threshold $\bm \mu$, a predicted moment is considered matched only if there is a corresponding ground truth with $\bm{\textbf{IoU} \geq \mu}$; otherwise, it receives a relevance score of 0. Once a ground truth moment is matched with a prediction, it is excluded from further matching. Please refer to~\cite{liang2024tvrrankingdatasetrankedvideo} for additional metric details.

%======================
\subsection{Implementation Details}
%======================

For segment retrieval, videos are divided into segments of $\tau_{S} = 4$ seconds, resulting in 383,828 segments. During online searching, the top $k = 200$ relevant segments are retrieved, with a Flat index used by default. For alternative index evaluations, we set the number of centroids to 8192 for both IVF and IVFPQ. In the case of IVFPQ, we set the number of sub-vectors for quantization to 16, allocating 8 bits per sub-vector. During search, we explore 128 clusters to retrieve the results for both IVF and IVFPQ.

To train the video and text feature projectors in \cref{sec:seg_retri}, we use CLIP ViT-L/14~\cite{radford2021learning} as the visual feature extractor, extracting frame features at 1 fps for segments, and its text encoder to extract query features, providing dual-modality inputs for both projectors. The visual feature projector includes a sequential Transformer with $L_{S}=6$ layers. The hidden and output sizes of both feature projectors are set to 768.  Training is conducted over 30 epochs with a learning rate of 0.0005, batch size of 256, using the AdamW optimizer~\cite{DBLP:conf/iclr/LoshchilovH19} with weight decay of 0.001 and gradient clipping of 5.0, and a cosine scheduler with warm-up.

For the refinement and re-ranking module in \cref{sec:ref-rerank-module}, we first pad each proposal with $\tau_{C}=8$ seconds on both sides to increase its likelihood of covering a matching moment. For the implementation using CLIP, we extract frame features at a rate of 1 fps from a proposal. For video and subtitle feature extraction utilizing I3D and RoBERTa, we segment a proposal into 1.5-second intervals, extracting features from each interval to form the feature sequence. We train both a small model with hidden size 384 and a large model with hidden size 768. All models are trained with a learning rate of 0.0001, batch size of 256, and other hyperparameters as in~\cite{DBLP:conf/sigir/0048SJNZZG21}. 
We have also explored various training objectives to optimize model performance for both implementations. The best-performing model using CLIP is achieved with the standard training setting described in~\cite{DBLP:conf/sigir/0048SJNZZG21}. The implementation with ReLoCLNet attains optimal results by incorporating two additional hinge losses: (i) between proposals containing ground truth moments and those not, and (ii) between strong and weak positive proposals to enforce their relative ranking. We applied a margin of 0.1 and a loss weight of 1 for these two hinge loss terms. Alternative training objectives are evaluated in ablation study in \cref{sec:ablation_ref_rerank}.

%======================
\begin{table}[t!]
    \centering
    \caption{\textbf{Performance on the TVR-Ranking validation and test sets using $\bm{\textbf{NDCG}@K,\textbf{IoU} \geq \mu}$ metrics.} Results of the three baselines are sourced from~\cite{liang2024tvrrankingdatasetrankedvideo}.  
    \textbf{SP}, \textbf{SPR$_{\text{CLIP}}$} and \textbf{SPR$_{\text{ReLo}}$} represent the results of coarse proposals and fine-grained results by two different instantiations in \cref{sec:inst-ref_rerank}}
    \resizebox{1.0\linewidth}{!}{
    \begin{tabular}{lccccccc}
        \toprule
        \multirow{2}{*}{\textbf{Model}}& \multicolumn{2}{c}{$\bm{\textbf{IoU}\geq 0.3}$} & \multicolumn{2}{c}{$\bm{\textbf{IoU}\geq 0.5}$} & \multicolumn{2}{c}{$\bm{\textbf{IoU}\geq 0.7}$} \\
        \cmidrule(lr){2-3} \cmidrule(lr){4-5} \cmidrule(lr){6-7}
         & \textbf{val} & \textbf{test} & \textbf{val} & \textbf{test} & \textbf{val} & \textbf{test} \\
        \midrule
        & \multicolumn{6}{c}{$\bm{\textbf{NDCG}@10}$} \\
        \cmidrule(lr){2-7}
        XML~\cite{lei2020tvr} & 0.2002 & 0.2044 & 0.1461 & 0.1502 & 0.0541 & 0.0589 \\
        CONQUER~\cite{DBLP:conf/mm/HouNC21} & 0.2450 & 0.2219 & 0.2262 & 0.2085 & 0.1670 & 0.1515 \\
        ReLoCLNet~\cite{DBLP:conf/sigir/0048SJNZZG21} & 0.4339 & 0.4353 & 0.3984 & 0.3986 & 0.2693 & 0.2807 \\
        SP & 0.4556 & 0.4713 & 0.3631 & 0.3646 & 0.2193 & 0.2236 \\
        SPR$_{\text{ReLo}}$ & \textbf{0.5373} & \textbf{0.5509} & \textbf{0.5084} & \textbf{0.5214} & \underline{0.3598} & \underline{0.3731} \\
        SPR$_{\text{CLIP}}$ & \underline{0.5139} & \underline{0.5162} & \underline{0.5061} & \underline{0.5079} & \textbf{0.4285} & \textbf{0.4305} \\
        \midrule
        & \multicolumn{6}{c}{$\bm{\textbf{NDCG}@20}$} \\
        \cmidrule(lr){2-7}
        XML & 0.2114 & 0.2167 & 0.1530 & 0.1590 & 0.0583 & 0.0635 \\
        %\midrule
        CONQUER & 0.2183 & 0.1968 & 0.2022 & 0.1851 & 0.1524 & 0.1365 \\
        %\midrule
        ReLoCLNet & 0.4418 & 0.4439 & 0.4060 & 0.4059 & 0.2787 & 0.2877 \\
        SP & 0.4510 & 0.4683 & 0.3580 & 0.3617 & 0.2142 & 0.2191 \\
        SPR$_{\text{ReLo}}$ & \textbf{0.5408} & \textbf{0.5523} & \textbf{0.5126} & \textbf{0.5231} & \underline{0.3590} & \underline{0.3718}\\
        SPR$_{\text{CLIP}}$ & \underline{0.5114} & \underline{0.5145} & \underline{0.5023} & \underline{0.5056} & \textbf{0.4249} & \textbf{0.4269} \\
        \midrule
        & \multicolumn{6}{c}{$\bm{\textbf{NDCG}@40}$} \\
        \cmidrule(lr){2-7}
        XML & 0.2408 & 0.2432 & 0.1740 & 0.1791 & 0.0666 & 0.0720 \\
        CONQUER & 0.2080 & 0.1885 & 0.1934 & 0.1775 & 0.1473 & 0.1323 \\
        ReLoCLNet & 0.4725 & 0.4735 & 0.4337 & 0.4337 & 0.3015 & 0.3079 \\
        SP & 0.4760 & 0.4910 & 0.3759 & 0.3774 & 0.2216 & 0.2254 \\
        SPR$_{\text{ReLo}}$ & \textbf{0.5642} & \textbf{0.5751} & \textbf{0.5310} & \textbf{0.5406} & \underline{0.3743} & \underline{0.3824} \\
        SPR$_{\text{CLIP}}$ & \underline{0.5329}  & \underline{0.5364} & \underline{0.5217} & \underline{0.5257} & \textbf{0.4415} & \textbf{0.4422} \\
        \bottomrule
    \end{tabular}
    }
    \label{tab:overall results}
\end{table}

%======================
\begin{table}[t!]
\centering
\caption{\textbf{Average processing time  (in second) for a single query at each stage.} SP is near real-time, and  SPR takes 0.7 to 1 second. All SPR implementations use Flat index for accuracy.}
\resizebox{1.0\linewidth}{!}{
\begin{tabular}{lcccc|c}
\toprule
\textbf{Model} & \textbf{Q. Emb.} & \textbf{Seg. Retr.} & \textbf{Prop. Gen.} & \textbf{Ref. \& Re-rank} & \textbf{Total} \\
\midrule
SP$_{\text{Flat}}$ & \multirow{6}{*}{$0.018$} & $0.126$ & \multirow{6}{*}{$0.0008$} & - & $0.144$ \\
SP$_{\text{IVF}}$ & ~ & $0.007$ & ~ & - & $0.026$ \\
SP$_{\text{IVFPQ}}$ & ~ & $0.004$ & ~ & - & $0.023$ \\
SPR$_{\text{ReLo}}$-S & ~ & $0.126$ & ~ & $0.775$ & $0.955$ \\
SPR$_{\text{ReLo}}$-L & ~ & $0.126$ & ~ & $0.781$ & $0.962$ \\
SPR$_{\text{CLIP}}$-L & ~ & $0.126$ & ~ & $0.498$ & $0.680$ \\
\bottomrule
\end{tabular}
}
\label{tab:single_q_infer}
\end{table}

%======================
\begin{table*}[t!]
\centering
\caption{\textbf{Scalability test} on the \textbf{T:} TVR-Ranking validation dataset additinal videos from \textbf{C:} Charades, and \textbf{A:} ActivityNet Captions. The NDCG values are computed for the coarse proposals (SP) over 500 validation queries in TVR-Ranking, each retrieving the top-200 segments. The last column reports the total time taken to process the 500 queries in parallel.  
}
\resizebox{1.0\linewidth}{!}{
\begin{tabular}{@{}lcccccccccccccc@{}}
\toprule
\multirow{2}{*}{\textbf{Corpus}} & \multirow{2}{*}{\textbf{\# Vid.}} & \multirow{2}{*}{\textbf{\# Seg.}} & \multirow{2}{*}{\textbf{Index}} & \multicolumn{3}{c}{$\bm{\textbf{NDCG}@10}$} & \multicolumn{3}{c}{$\bm{\textbf{NDCG}@20}$} & \multicolumn{3}{c}{$\bm{\textbf{NDCG}@40}$} & \multirow{2}{*}{\textbf{Retr. Time}} \\ 
\cmidrule(lr){5-7} \cmidrule(lr){8-10} \cmidrule(lr){11-13}
 & & & & $\bm{\textbf{IoU}\geq 0.3}$ & $\bm{\textbf{IoU}\geq 0.5}$ & $\bm{\textbf{IoU}\geq 0.7}$ & $\bm{\textbf{IoU}\geq 0.3}$ & $\bm{\textbf{IoU}\geq 0.5}$ & $\bm{\textbf{IoU}\geq 0.7}$ & $\bm{\textbf{IoU}\geq 0.3}$ & $\bm{\textbf{IoU}\geq 0.5}$ & $\bm{\textbf{IoU}\geq 0.7}$ & \\
\midrule
\multirow{3}{*}{T} & \multirow{3}{*}{19,614} & \multirow{3}{*}{383,828} & Flat & 0.4556 & 0.3631 & 0.2193 & 0.4510 & 0.3580 & 0.2142 & 0.4760 & 0.3759 & 0.2216 & 0.74 \\
&  &  & IVF & 0.4385 & 0.3460 & 0.2040 & 0.4316 & 0.3397 & 0.1984 & 0.4490 & 0.3521 & 0.2028 & 0.58 \\
&  &  & IVFPQ & 0.0920 & 0.0731 & 0.0495 & 0.0930 & 0.0724 & 0.0484 & 0.1013 & 0.0775 & 0.0497 & 0.29 \\ 
%\cmidrule(lr){4-14}
\cmidrule{1-14}
\multirow{3}{*}{T+C} & \multirow{3}{*}{29,462} & \multirow{3}{*}{460,443} & Flat & 0.4557 & 0.3635 & 0.2192 & 0.4504 & 0.3579 & 0.2137 & 0.4740 & 0.3749 & 0.2204 & 0.90 \\
 & &  & IVF & 0.4384 & 0.3469 & 0.2068 & 0.4314 & 0.3387 & 0.1990 & 0.4489 & 0.3519 & 0.2045 & 0.66 \\
 & &  & IVFPQ & 0.0814 & 0.0659 & 0.0467 & 0.0835 & 0.0647 & 0.0427 & 0.0908 & 0.0686 & 0.0432 & 0.29 \\
%\cmidrule{4-14}
\cmidrule{1-14}
\multirow{3}{*}{T+A} & \multirow{3}{*}{33,087} & \multirow{3}{*}{784,302} & Flat & 0.4548 & 0.3629 & 0.2177 & 0.4486 & 0.3564 & 0.2118 & 0.4721 & 0.3733 & 0.2186 & 1.33 \\
 & &  & IVF & 0.4288 & 0.3332 & 0.1970 & 0.4212 & 0.3259 & 0.1902 & 0.4396 & 0.3396 & 0.1958 & 0.85 \\
 &  &  & IVFPQ & 0.0695 & 0.0551 & 0.0399 & 0.0711 & 0.0568 & 0.0391 & 0.0774 & 0.0602 & 0.0397 & 0.23 \\ 
%\cmidrule{4-14}
\cmidrule{1-14}
\multirow{3}{*}{T+C+A} & \multirow{3}{*}{42,935} & \multirow{3}{*}{860,917} & Flat & 0.4551 & 0.3617 & 0.2167 & 0.4483 & 0.3547 & 0.2106 & 0.4709 & 0.3710 & 0.2170 & 1.50 \\
 & &  & IVF & 0.4367 & 0.3456 & 0.2005 & 0.4285 & 0.3372 & 0.1930 & 0.4473 & 0.3505 & 0.1980 & 0.90 \\
 &  &  & IVFPQ & 0.0695 & 0.0576 & 0.0441 & 0.0735 & 0.0584 & 0.0428 & 0.0789 & 0.0616 & 0.0427 & 0.23 \\
\bottomrule
\end{tabular}
}

\label{tab:seg_retri_scalability}
\end{table*}

%======================
\subsection{Performance Overview}
%======================

The performances of the proposed frameworks, SP and SPR, alongside the RVMR baseline models are presented in \cref{tab:overall results}. Here, SP represents coarse results obtained through the Segment-Proposal process, while SPR$_{\text{CLIP}}$ and SPR$_{\text{ReLo}}$ denote two different instantiations used for refinement and re-ranking (see \cref{sec:inst-ref_rerank}). Notably, SPR$_{\text{ReLo}}$ utilizes the same features as the baseline models for direct comparison. All baseline model results are sourced from \cite{liang2024tvrrankingdatasetrankedvideo}, with training conditions matched to ours by using the top 40 moments for each query in the pseudo training set.

Observe that the coarse moment results from the Segment-Proposal (SP) process already surpass ReLoCLNet, the best baseline, across all $\text{NDCG}@N,\text{IoU}\geq 0.3$ metrics. However, as the IoU threshold increases, SP performs slightly worse than ReLoCLNet. Although efficient, SP lacks temporal precision because a proposal's duration is fixed by the number of segments in the proposal. Additionally, the proposal ranking may not be optimal.

With refinement and re-ranking, both SPR instantiations demonstrate significant performance improvements across all metrics, outperforming other methods by a substantial margin. Notably, SPR$_{\text{ReLo}}$ performs better at lower IoU thresholds, while SPR$_{\text{CLIP}}$ excels on metrics with $\text{IoU} \geq 0.7$. SPR$_{\text{ReLo}}$ may benefit from richer information during proposal ranking by leveraging both video and subtitle features for video modeling. In contrast, the shorter minimum time scale of the visual CLIP features (1 vs. 1.5 seconds) enhances SPR$_{\text{CLIP}}$’s ability to predict precise timestamps.

%======================
\subsection{Efficiency Test}

We simulate a real-world scenario in which a user submits a text query and waits online for the relevant moments to be retrieved. We record the processing time at each stage and report the average time cost across multiple queries, with query lengths ranging from 5 to 50 words (see \cref{tab:single_q_infer}). Experiments are conducted using an NVIDIA Tesla V100 GPU and an Intel Xeon Gold 6150 CPU.

We evaluate SP across three indexes: Flat, IVF, and IVFPQ; all of which achieve near real-time response. The Flat index, which provides the highest quality segments, takes 0.126 seconds, while the IVF and IVFPQ indexes require less than 0.01 seconds. Additional refinement and re-ranking take approximately 0.5 to 0.8 seconds for both implementations. The suffixes `-S' and `-L' after the model name denote the small (hidden size 384) and large (hidden size 768) models evaluated. In short, the total inference time for SPR to process a query is about 0.7–1 second.

%======================
\begin{table*}[t!]
\centering
\caption{\textbf{Upper bound performance of the SPR pipeline} on the TVR-Ranking validation set, based on coarse proposals generated from the top-200 segments. $\tau_{C}$ represents the context length padded to each proposal. The minimum time scale is determined by the frame sampling rate used for feature extraction}
\resizebox{1.0\linewidth}{!}{
\begin{tabular}{@{}lccccccccccc@{}}
\toprule
\multirow{2}{*}{\textbf{Group}} & \multirow{2}{*}{$\bm{\tau_{C}}$} & \multirow{2}{*}{\textbf{Min. Time Scale}} & \multicolumn{3}{c}{$\bm{\textbf{NDCG}@10}$} & \multicolumn{3}{c}{$\bm{\textbf{NDCG}@20}$} & \multicolumn{3}{c}{$\bm{\textbf{NDCG}@40}$} \\ 
\cmidrule(lr){4-6} \cmidrule(lr){7-9} \cmidrule(lr){10-12}
 & & & $\bm{\textbf{IoU}\geq 0.3}$ & $\bm{\textbf{IoU}\geq 0.5}$ & $\bm{\textbf{IoU}\geq 0.7}$ & $\bm{\textbf{IoU}\geq 0.3}$ & $\bm{\textbf{IoU}\geq 0.5}$ & $\bm{\textbf{IoU}\geq 0.7}$ & $\bm{\textbf{IoU}\geq 0.3}$ & $\bm{\textbf{IoU}\geq 0.5}$ & $\bm{\textbf{IoU}\geq 0.7}$ \\
\midrule
SP & - & 4 & 0.4556 & 0.3631 & 0.2193 & 0.4510 & 0.3580 & 0.2142 & 0.4760 & 0.3759 & 0.2216 \\
UB & 0 & - & 0.8694 & 0.8563 & 0.8298 & 0.8203 & 0.8040 & 0.7705 & 0.7827 & 0.7652 & 0.7289 \\
UB       & 4 & - & 0.8883 & 0.8832 & 0.8748 & 0.8409 & 0.8346 & 0.8250 & 0.8040 & 0.7971 & 0.7869 \\
UB       & 8 & - & 0.8909 & 0.8880 & 0.8842 & 0.8443 & 0.8407 & 0.8356 & 0.8077 & 0.8037 & 0.7982 \\
PUB & 8 & 1 & 0.8837 & 0.8807 & 0.8686 & 0.8373 & 0.8337 & 0.8169 & 0.8009 & 0.7969 & 0.7776 \\
PUB & 8 & 1.5 & 0.8847 & 0.8780 & 0.8418 & 0.8384 & 0.8299 & 0.7814 & 0.8021 & 0.7927 & 0.7378 \\
\bottomrule
\end{tabular}
}
\label{tab:practical_upper_bound}
\end{table*}

%======================
\begin{table*}[t!]
\centering
\caption{\textbf{Ablation study on training strategies}  refinement and re-ranking. `S' and `L' refer to model size; `HN' is the use of hard negative samples for a query. `++' and `+' refer to the number of strong and weak positive training moments for a query, respectively.}
\resizebox{1.0\linewidth}{!}{
\begin{tabular}{lcccccccccc}
\toprule
\multirow{2}{*}{\textbf{Model}} & \multirow{2}{*}{\textbf{Group}} & \multicolumn{3}{c}{$\bm{\textbf{NDCG}@10}$} & \multicolumn{3}{c}{$\bm{\textbf{NDCG}@20}$} & \multicolumn{3}{c}{$\bm{\textbf{NDCG}@40}$} \\ 
\cmidrule(lr){3-5} \cmidrule(lr){6-8} \cmidrule(lr){9-11}
 & & $\bm{\textbf{IoU}\geq 0.3}$ & $\bm{\textbf{IoU}\geq 0.5}$ & $\bm{\textbf{IoU}\geq 0.7}$ & $\bm{\textbf{IoU}\geq 0.3}$ & $\bm{\textbf{IoU}\geq 0.5}$ & $\bm{\textbf{IoU}\geq 0.7}$ & $\bm{\textbf{IoU}\geq 0.3}$ & $\bm{\textbf{IoU}\geq 0.5}$ & $\bm{\textbf{IoU}\geq 0.7}$ \\
\midrule
SP & - & 0.4556 & 0.3631 & 0.2193 & 0.4510 & 0.3580 & 0.2142 & 0.4760 & 0.3759 & 0.2216 \\
\midrule
\multirow{5}{*}{$\text{SPR}_{\text{ReLo}}$-S} & Standard & 0.4226 & 0.3938 & 0.2699 & 0.4365 & 0.4065 & 0.2779 & 0.4693 & 0.4354 & 0.2973 \\
 & Hard Neg. (HN) & 0.4338 & 0.4114 & 0.2909 & 0.4497 & 0.4246 & 0.3003 & 0.4863 & 0.4573 & 0.3207 \\
 & HN w. 5++,20+ & 0.4532 & 0.4278 & 0.2902 & 0.4642 & 0.4362 & 0.2966 & 0.4951 & 0.4632 & 0.3154 \\
 & HN w. 5++,30+ & 0.4573 & 0.4304 & 0.2974 & 0.4694 & 0.4394 & 0.3041 & 0.4996 & 0.4658 & 0.3220 \\
 & HN w. 10++,20+ & 0.4620 & 0.4361 & 0.2969 & 0.4734 & 0.4463 & 0.3055 & 0.5038 & 0.4726 & 0.3230 \\
\midrule
\multirow{5}{*}{$\text{SPR}_{\text{ReLo}}$-L} & Standard & 0.5309 & 0.5038 & 0.3555 & 0.5353 & 0.5084 & 0.3593 & 0.5567 & 0.5276 & 0.3727 \\
 & Hard Neg. (HN) & 0.5288 & 0.5005 & 0.3540 & 0.5341 & 0.5048 & 0.3567 & 0.5566 & 0.5248 & 0.3709 \\
 & HN w. 5++,20+ & 0.5190 & 0.4896 & 0.3448 & 0.5263 & 0.4969 & 0.3499 & 0.5482 & 0.5164 & 0.3632 \\
 & HN w. 5++,30+ & 0.5328 & 0.5057 & 0.3591 & 0.5361 & 0.5083 & 0.3580 & 0.5603 & 0.5294 & 0.3738 \\
 & HN w. 10++,20+ & 0.5373 & 0.5084 & 0.3598 & 0.5408 & 0.5126 & 0.3590 & 0.5642 & 0.5310 & 0.3743 \\
\midrule
\multirow{3}{*}{$\text{SPR}_{\text{CLIP}}$-L} & Standard & 0.5139 & 0.5061 & 0.4285 & 0.5114 & 0.5023 & 0.4249 & 0.5329 & 0.5217 & 0.4415 \\
 & Hard Neg. (HN) & 0.4970 & 0.4887 & 0.4146 & 0.4988 & 0.4892 & 0.4137 & 0.5229 & 0.5110 & 0.4305 \\
 & HN w. 5++,20+ & 0.4645 & 0.4558 & 0.3839 & 0.4712 & 0.4606 & 0.3874 & 0.4980 & 0.4842 & 0.4065 \\
\bottomrule
\end{tabular}
}
\label{tab:refinement_result}
\end{table*}
%======================

%====================================
\subsection{Scalability Test}
%====================================

We conduct a scalability test by expanding the search space with a large number of unrelated videos. Specifically, we added videos from two other datasets, Charades~\cite{DBLP:conf/eccv/SigurdssonVWFLG16} and ActivityNet Captions~\cite{DBLP:conf/iccv/KrishnaHRFN17}, to the TVR-Ranking dataset. Charades includes 9,848 videos focused on indoor human activities, while ActivityNet Captions contains 13,473 YouTube videos. Since these videos belong to a different domain than TVR-Ranking, we assume they contain no matching moments for TVR-Ranking queries.

Given that refinement \& re-ranking relies solely on the top-$k$ retrieved segments, independent of the video collection size, we assess scalability using the coarse proposal results with the top-200 segments. \cref{tab:seg_retri_scalability} presents the SP performance and overall retrieval time using the 500 validation queries from TVR-Ranking on both the original and the expanded corpus.

Across all NDCG metrics, we observe minimal performance degradation with the Flat and IVF indexes as the corpus size expands with the addition of irrelevant videos, increasing from 0.38 million to 0.86 million segments. While IVFPQ is the fastest index, it consistently performs the worst in our setting. The primary reason is the lack of information redundancy in the query and segment representations. Similar observations are made in other systems, such as
 DPR~\cite{DBLP:conf/emnlp/KarpukhinOMLWEC20}, ANCE~\cite{DBLP:conf/iclr/XiongXLTLBAO21}, and STAR~\cite{DBLP:conf/sigir/ZhanM0G0M21}, which also use Flat indexes during inference to preserve performance.

The last column in \cref{tab:seg_retri_scalability} is the retrieval time used by processing all the 500 validation queries. All indexes are extremely fast and all queries are processed in parallel. We find that the time cost for segment retrieval increases linearly with the Flat index and sub-linearly with IVF index.

Overall, our results confirm that the SP framework is both robust and efficient, demonstrating good scalability.

%===============================================
\subsection{Upper bound Analysis on SPR Framework}
%===============================================
Since refinement and re-ranking depend solely on the top-$k$ retrieved segments, the quality of these segment sets establishes an upper bound for the final performance. An ideal refinement module would predict the perfect timestamps for any ground truth moments fully or partially contained within a proposal. The ideal re-ranking module would then provide an optimal ranking based on a specified IoU threshold $\mu$. The NDCG score of this ranked list provides a theoretical upper bound based on the current coarse proposals.
We denote these values as ``UB'' in \cref{tab:practical_upper_bound}. However, due to the design constraints of our refinement and re-ranking model, timestamp prediction is not continuous but instead a multiple of the minimum time scale (depends on frame sampling rate) of the video feature sequence. To account for this, we calculate a practical upper bound under this constraint, denoted as ``PUB'' in the table. 

Results in \cref{tab:practical_upper_bound} showing that the upper bound of proposals without the added padding context (\ie $\bm{\tau_{C}}$=0) is already high, and reaches a peak with a context length of 8 seconds. The high upper bounds offer promising prospects for optimal model design, which validates the rationale behind SPR. Compared to the practical upper bound, there are room for further improvement for refinement and re-ranking. However, there is also a trade-off between performance and inference efficiency. The gap between the current and optimal results may also be attributed to the quality of the training data, which is not manually annotated.

%===============================================
\subsection{Ablation Study on Training Strategies}
\label{sec:ablation_ref_rerank}
%===============================================

For the refinement and re-ranking module, on top of the default training setting, we experiment with variations in two areas. First, we evaluate the impact of model size by adjusting the hidden layer size, with the suffixes ``-S'' and ``-L'' denoting small and large models.

We also experiment with additional loss terms. Building on the default training objective (referred to as `Standard'), we first introduce coarse proposals that do not contain any ground truth as hard negative (`HN') samples. By applying a video-level hinge loss between positive samples and these hard negatives, the model learns to prioritize coarse proposals that contain ground truth. Additionally, to ensure that positive samples with higher relevance are ranked above those with lower relevance, we split the training samples into strong and weak positives. Specifically, in the pseudo-training set, which has $N$ positive moments for each query, we select the top $N_{s}$ samples as strong positives and the last $N_{w}$ samples as weak positives, where $N_{s} + N_{w} \leq N$. For each query in a batch, a strong positive and a weak positive are randomly sampled.

As reported in \cref{tab:refinement_result}, for $\text{SPR}_{\text{ReLo}}$-S, incorporating hard negatives positively impacts performance, and adding a relative ranking constraint further improves results. In contrast, for larger models like $\text{SPR}_{\text{ReLo}}$-L and $\text{SPR}_{\text{CLIP}}$-L, training with the `Standard' setting yields nearly the best performance. We speculate that larger models can effectively learns from standard positive-negative pairs when trained with sufficient samples. For the $\text{SPR}_{\text{CLIP}}$-L model, the `Standard' setting outperforms the rest. This is because the training samples ranked between $N_{s}$ and $N-N_{w}$ are excluded in training. The sample reduction outweighs the benefits of the explicit ranking-related constraints.

%======================
\section{Conclusion}
\label{sec:conclusion}
%======================
In this paper, we propose a video moment search framework designed for real-world applications. We address limitations in existing solutions, such as the unrealistic assumption of a single `perfect match' moment per query, challenges in processing hour-long videos, and inefficiencies in inference. Our framework decomposes the moment search process into three independent stages and standardizes video processing with a fixed unit of segments. Additionally, we incorporate techniques from dense retrieval to improve efficiency and scalability. Our framework achieves state-of-the-art performance on the TVR-Ranking dataset, significantly reducing computational costs and processing time. Additionally, the scalability of our approach opens up possibilities for large-scale deployment. Future work could explore leveraging more advanced neural architectures and fine-tuning techniques to further improve the three independent stages.

\bibliographystyle{ACM-Reference-Format}
\bibliography{sample-base}


\begin{thebibliography}{63}

%%% the Web address.

\ifx \showCODEN    \undefined \def \showCODEN     #1{\unskip}     \fi
\ifx \showDOI      \undefined \def \showDOI       #1{#1}\fi
\ifx \showISBNx    \undefined \def \showISBNx     #1{\unskip}     \fi
\ifx \showISBNxiii \undefined \def \showISBNxiii  #1{\unskip}     \fi
\ifx \showISSN     \undefined \def \showISSN      #1{\unskip}     \fi
\ifx \showLCCN     \undefined \def \showLCCN      #1{\unskip}     \fi
\ifx \shownote     \undefined \def \shownote      #1{#1}          \fi
\ifx \showarticletitle \undefined \def \showarticletitle #1{#1}   \fi
\ifx \showURL      \undefined \def \showURL       {\relax}        \fi
% invisible to TeX
\providecommand\bibfield[2]{#2}
\providecommand\bibinfo[2]{#2}
\providecommand\natexlab[1]{#1}
\providecommand\showeprint[2][]{arXiv:#2}

\bibitem[Barrios et~al\mbox{.}(2023)]%
        {DBLP:conf/iccv/BarriosSCHG23}
\bibfield{author}{\bibinfo{person}{Wayner Barrios}, \bibinfo{person}{Mattia Soldan}, \bibinfo{person}{Alberto~Mario Ceballos{-}Arroyo}, \bibinfo{person}{Fabian~Caba Heilbron}, {and} \bibinfo{person}{Bernard Ghanem}.} \bibinfo{year}{2023}\natexlab{}.
\newblock \showarticletitle{Localizing Moments in Long Video Via Multimodal Guidance}. In \bibinfo{booktitle}{\emph{{IEEE/CVF} International Conference on Computer Vision, {ICCV} 2023, Paris, France, October 1-6, 2023}}. \bibinfo{publisher}{{IEEE}}, \bibinfo{pages}{13621--13632}.
\newblock
\urldef\tempurl%
\url{https://doi.org/10.1109/ICCV51070.2023.01257}
\showDOI{\tempurl}


\bibitem[Bendersky et~al\mbox{.}(2011)]%
        {DBLP:conf/wsdm/BenderskyCD11}
\bibfield{author}{\bibinfo{person}{Michael Bendersky}, \bibinfo{person}{W.~Bruce Croft}, {and} \bibinfo{person}{Yanlei Diao}.} \bibinfo{year}{2011}\natexlab{}.
\newblock \showarticletitle{Quality-biased ranking of web documents}. In \bibinfo{booktitle}{\emph{Proceedings of the Forth International Conference on Web Search and Web Data Mining, {WSDM} 2011, Hong Kong, China, February 9-12, 2011}}, \bibfield{editor}{\bibinfo{person}{Irwin King}, \bibinfo{person}{Wolfgang Nejdl}, {and} \bibinfo{person}{Hang Li}} (Eds.). \bibinfo{publisher}{{ACM}}, \bibinfo{pages}{95--104}.
\newblock
\urldef\tempurl%
\url{https://doi.org/10.1145/1935826.1935849}
\showDOI{\tempurl}


\bibitem[Borgeaud et~al\mbox{.}(2022)]%
        {DBLP:conf/icml/BorgeaudMHCRM0L22}
\bibfield{author}{\bibinfo{person}{Sebastian Borgeaud}, \bibinfo{person}{Arthur Mensch}, \bibinfo{person}{Jordan Hoffmann}, \bibinfo{person}{Trevor Cai}, \bibinfo{person}{Eliza Rutherford}, \bibinfo{person}{Katie Millican}, \bibinfo{person}{George van~den Driessche}, \bibinfo{person}{Jean{-}Baptiste Lespiau}, \bibinfo{person}{Bogdan Damoc}, \bibinfo{person}{Aidan Clark}, \bibinfo{person}{Diego de Las~Casas}, \bibinfo{person}{Aurelia Guy}, \bibinfo{person}{Jacob Menick}, \bibinfo{person}{Roman Ring}, \bibinfo{person}{Tom Hennigan}, \bibinfo{person}{Saffron Huang}, \bibinfo{person}{Loren Maggiore}, \bibinfo{person}{Chris Jones}, \bibinfo{person}{Albin Cassirer}, \bibinfo{person}{Andy Brock}, \bibinfo{person}{Michela Paganini}, \bibinfo{person}{Geoffrey Irving}, \bibinfo{person}{Oriol Vinyals}, \bibinfo{person}{Simon Osindero}, \bibinfo{person}{Karen Simonyan}, \bibinfo{person}{Jack~W. Rae}, \bibinfo{person}{Erich Elsen}, {and} \bibinfo{person}{Laurent Sifre}.} \bibinfo{year}{2022}\natexlab{}.
\newblock \showarticletitle{Improving Language Models by Retrieving from Trillions of Tokens}. In \bibinfo{booktitle}{\emph{International Conference on Machine Learning, {ICML} 2022, 17-23 July 2022, Baltimore, Maryland, {USA}}} \emph{(\bibinfo{series}{Proceedings of Machine Learning Research}, Vol.~\bibinfo{volume}{162})}, \bibfield{editor}{\bibinfo{person}{Kamalika Chaudhuri}, \bibinfo{person}{Stefanie Jegelka}, \bibinfo{person}{Le~Song}, \bibinfo{person}{Csaba Szepesv{\'{a}}ri}, \bibinfo{person}{Gang Niu}, {and} \bibinfo{person}{Sivan Sabato}} (Eds.). \bibinfo{publisher}{{PMLR}}, \bibinfo{pages}{2206--2240}.
\newblock
\urldef\tempurl%
\url{https://proceedings.mlr.press/v162/borgeaud22a.html}
\showURL{%
\tempurl}


\bibitem[Burges et~al\mbox{.}(2005)]%
        {10.1145/1102351.1102363}
\bibfield{author}{\bibinfo{person}{Chris Burges}, \bibinfo{person}{Tal Shaked}, \bibinfo{person}{Erin Renshaw}, \bibinfo{person}{Ari Lazier}, \bibinfo{person}{Matt Deeds}, \bibinfo{person}{Nicole Hamilton}, {and} \bibinfo{person}{Greg Hullender}.} \bibinfo{year}{2005}\natexlab{}.
\newblock \showarticletitle{Learning to rank using gradient descent}. In \bibinfo{booktitle}{\emph{Proceedings of the 22nd International Conference on Machine Learning}} (Bonn, Germany) \emph{(\bibinfo{series}{ICML '05})}. \bibinfo{publisher}{Association for Computing Machinery}, \bibinfo{address}{New York, NY, USA}, \bibinfo{pages}{89–96}.
\newblock
\showISBNx{1595931805}
\urldef\tempurl%
\url{https://doi.org/10.1145/1102351.1102363}
\showDOI{\tempurl}


\bibitem[Cao et~al\mbox{.}(2021)]%
        {DBLP:conf/emnlp/CaoCSZZ21}
\bibfield{author}{\bibinfo{person}{Meng Cao}, \bibinfo{person}{Long Chen}, \bibinfo{person}{Mike~Zheng Shou}, \bibinfo{person}{Can Zhang}, {and} \bibinfo{person}{Yuexian Zou}.} \bibinfo{year}{2021}\natexlab{}.
\newblock \showarticletitle{On Pursuit of Designing Multi-modal Transformer for Video Grounding}. In \bibinfo{booktitle}{\emph{Proceedings of the 2021 Conference on Empirical Methods in Natural Language Processing, {EMNLP} 2021, Virtual Event / Punta Cana, Dominican Republic, 7-11 November, 2021}}, \bibfield{editor}{\bibinfo{person}{Marie{-}Francine Moens}, \bibinfo{person}{Xuanjing Huang}, \bibinfo{person}{Lucia Specia}, {and} \bibinfo{person}{Scott~Wen{-}tau Yih}} (Eds.). \bibinfo{publisher}{Association for Computational Linguistics}, \bibinfo{pages}{9810--9823}.
\newblock
\urldef\tempurl%
\url{https://doi.org/10.18653/V1/2021.EMNLP-MAIN.773}
\showDOI{\tempurl}


\bibitem[Carreira and Zisserman(2017)]%
        {carreira2017quo}
\bibfield{author}{\bibinfo{person}{Joao Carreira} {and} \bibinfo{person}{Andrew Zisserman}.} \bibinfo{year}{2017}\natexlab{}.
\newblock \showarticletitle{Quo vadis, action recognition? a new model and the kinetics dataset}. In \bibinfo{booktitle}{\emph{proceedings of the IEEE Conference on Computer Vision and Pattern Recognition}}. \bibinfo{pages}{6299--6308}.
\newblock


\bibitem[Chen et~al\mbox{.}(2023)]%
        {DBLP:conf/iclr/ChenHSC23}
\bibfield{author}{\bibinfo{person}{Wenhu Chen}, \bibinfo{person}{Hexiang Hu}, \bibinfo{person}{Chitwan Saharia}, {and} \bibinfo{person}{William~W. Cohen}.} \bibinfo{year}{2023}\natexlab{}.
\newblock \showarticletitle{Re-Imagen: Retrieval-Augmented Text-to-Image Generator}. In \bibinfo{booktitle}{\emph{The Eleventh International Conference on Learning Representations, {ICLR} 2023, Kigali, Rwanda, May 1-5, 2023}}. \bibinfo{publisher}{OpenReview.net}.
\newblock
\urldef\tempurl%
\url{https://openreview.net/forum?id=XSEBx0iSjFQ}
\showURL{%
\tempurl}


\bibitem[Cheng et~al\mbox{.}(2023)]%
        {DBLP:conf/cvpr/ChengW0CBB23}
\bibfield{author}{\bibinfo{person}{Feng Cheng}, \bibinfo{person}{Xizi Wang}, \bibinfo{person}{Jie Lei}, \bibinfo{person}{David~J. Crandall}, \bibinfo{person}{Mohit Bansal}, {and} \bibinfo{person}{Gedas Bertasius}.} \bibinfo{year}{2023}\natexlab{}.
\newblock \showarticletitle{VindLU: {A} Recipe for Effective Video-and-Language Pretraining}. In \bibinfo{booktitle}{\emph{{IEEE/CVF} Conference on Computer Vision and Pattern Recognition, {CVPR} 2023, Vancouver, BC, Canada, June 17-24, 2023}}. \bibinfo{publisher}{{IEEE}}, \bibinfo{pages}{10739--10750}.
\newblock
\urldef\tempurl%
\url{https://doi.org/10.1109/CVPR52729.2023.01034}
\showDOI{\tempurl}


\bibitem[Datta et~al\mbox{.}(2008)]%
        {DBLP:journals/csur/DattaJLW08}
\bibfield{author}{\bibinfo{person}{Ritendra Datta}, \bibinfo{person}{Dhiraj Joshi}, \bibinfo{person}{Jia Li}, {and} \bibinfo{person}{James~Ze Wang}.} \bibinfo{year}{2008}\natexlab{}.
\newblock \showarticletitle{Image retrieval: Ideas, influences, and trends of the new age}.
\newblock \bibinfo{journal}{\emph{{ACM} Comput. Surv.}} \bibinfo{volume}{40}, \bibinfo{number}{2} (\bibinfo{year}{2008}), \bibinfo{pages}{5:1--5:60}.
\newblock
\urldef\tempurl%
\url{https://doi.org/10.1145/1348246.1348248}
\showDOI{\tempurl}


\bibitem[Devlin(2018)]%
        {devlin2018bert}
\bibfield{author}{\bibinfo{person}{Jacob Devlin}.} \bibinfo{year}{2018}\natexlab{}.
\newblock \showarticletitle{Bert: Pre-training of deep bidirectional transformers for language understanding}.
\newblock \bibinfo{journal}{\emph{arXiv preprint arXiv:1810.04805}} (\bibinfo{year}{2018}).
\newblock


\bibitem[Douze et~al\mbox{.}(2024)]%
        {DBLP:journals/corr/abs-2401-08281}
\bibfield{author}{\bibinfo{person}{Matthijs Douze}, \bibinfo{person}{Alexandr Guzhva}, \bibinfo{person}{Chengqi Deng}, \bibinfo{person}{Jeff Johnson}, \bibinfo{person}{Gergely Szilvasy}, \bibinfo{person}{Pierre{-}Emmanuel Mazar{\'{e}}}, \bibinfo{person}{Maria Lomeli}, \bibinfo{person}{Lucas Hosseini}, {and} \bibinfo{person}{Herv{\'{e}} J{\'{e}}gou}.} \bibinfo{year}{2024}\natexlab{}.
\newblock \showarticletitle{The Faiss library}.
\newblock \bibinfo{journal}{\emph{CoRR}}  \bibinfo{volume}{abs/2401.08281} (\bibinfo{year}{2024}).
\newblock
\urldef\tempurl%
\url{https://doi.org/10.48550/ARXIV.2401.08281}
\showDOI{\tempurl}
\showeprint[arXiv]{2401.08281}


\bibitem[Escorcia et~al\mbox{.}(2019)]%
        {DBLP:journals/corr/abs-1907-12763}
\bibfield{author}{\bibinfo{person}{Victor Escorcia}, \bibinfo{person}{Mattia Soldan}, \bibinfo{person}{Josef Sivic}, \bibinfo{person}{Bernard Ghanem}, {and} \bibinfo{person}{Bryan~C. Russell}.} \bibinfo{year}{2019}\natexlab{}.
\newblock \showarticletitle{Temporal Localization of Moments in Video Collections with Natural Language}.
\newblock \bibinfo{journal}{\emph{CoRR}}  \bibinfo{volume}{abs/1907.12763} (\bibinfo{year}{2019}).
\newblock
\showeprint[arXiv]{1907.12763}
\urldef\tempurl%
\url{http://arxiv.org/abs/1907.12763}
\showURL{%
\tempurl}


\bibitem[Gao et~al\mbox{.}(2017)]%
        {DBLP:conf/iccv/GaoSYN17}
\bibfield{author}{\bibinfo{person}{Jiyang Gao}, \bibinfo{person}{Chen Sun}, \bibinfo{person}{Zhenheng Yang}, {and} \bibinfo{person}{Ram Nevatia}.} \bibinfo{year}{2017}\natexlab{}.
\newblock \showarticletitle{{TALL:} Temporal Activity Localization via Language Query}. In \bibinfo{booktitle}{\emph{{IEEE} International Conference on Computer Vision, {ICCV} 2017, Venice, Italy, October 22-29, 2017}}. \bibinfo{publisher}{{IEEE} Computer Society}, \bibinfo{pages}{5277--5285}.
\newblock
\urldef\tempurl%
\url{https://doi.org/10.1109/ICCV.2017.563}
\showDOI{\tempurl}


\bibitem[Gao et~al\mbox{.}(2021a)]%
        {DBLP:conf/naacl/GaoDC21}
\bibfield{author}{\bibinfo{person}{Luyu Gao}, \bibinfo{person}{Zhuyun Dai}, {and} \bibinfo{person}{Jamie Callan}.} \bibinfo{year}{2021}\natexlab{a}.
\newblock \showarticletitle{{COIL:} Revisit Exact Lexical Match in Information Retrieval with Contextualized Inverted List}. In \bibinfo{booktitle}{\emph{Proceedings of the 2021 Conference of the North American Chapter of the Association for Computational Linguistics: Human Language Technologies, {NAACL-HLT} 2021, Online, June 6-11, 2021}}, \bibfield{editor}{\bibinfo{person}{Kristina Toutanova}, \bibinfo{person}{Anna Rumshisky}, \bibinfo{person}{Luke Zettlemoyer}, \bibinfo{person}{Dilek Hakkani{-}T{\"{u}}r}, \bibinfo{person}{Iz~Beltagy}, \bibinfo{person}{Steven Bethard}, \bibinfo{person}{Ryan Cotterell}, \bibinfo{person}{Tanmoy Chakraborty}, {and} \bibinfo{person}{Yichao Zhou}} (Eds.). \bibinfo{publisher}{Association for Computational Linguistics}, \bibinfo{pages}{3030--3042}.
\newblock
\urldef\tempurl%
\url{https://doi.org/10.18653/V1/2021.NAACL-MAIN.241}
\showDOI{\tempurl}


\bibitem[Gao et~al\mbox{.}(2021b)]%
        {DBLP:conf/emnlp/GaoYC21}
\bibfield{author}{\bibinfo{person}{Tianyu Gao}, \bibinfo{person}{Xingcheng Yao}, {and} \bibinfo{person}{Danqi Chen}.} \bibinfo{year}{2021}\natexlab{b}.
\newblock \showarticletitle{SimCSE: Simple Contrastive Learning of Sentence Embeddings}. In \bibinfo{booktitle}{\emph{Proceedings of the 2021 Conference on Empirical Methods in Natural Language Processing, {EMNLP} 2021, Virtual Event / Punta Cana, Dominican Republic, 7-11 November, 2021}}, \bibfield{editor}{\bibinfo{person}{Marie{-}Francine Moens}, \bibinfo{person}{Xuanjing Huang}, \bibinfo{person}{Lucia Specia}, {and} \bibinfo{person}{Scott~Wen{-}tau Yih}} (Eds.). \bibinfo{publisher}{Association for Computational Linguistics}, \bibinfo{pages}{6894--6910}.
\newblock
\urldef\tempurl%
\url{https://doi.org/10.18653/V1/2021.EMNLP-MAIN.552}
\showDOI{\tempurl}


\bibitem[Gupta et~al\mbox{.}(2024)]%
        {DBLP:conf/coling/GuptaAD24}
\bibfield{author}{\bibinfo{person}{Deepak Gupta}, \bibinfo{person}{Kush Attal}, {and} \bibinfo{person}{Dina Demner{-}Fushman}.} \bibinfo{year}{2024}\natexlab{}.
\newblock \showarticletitle{Towards Answering Health-related Questions from Medical Videos: Datasets and Approaches}. In \bibinfo{booktitle}{\emph{Proceedings of the 2024 Joint International Conference on Computational Linguistics, Language Resources and Evaluation, {LREC/COLING} 2024, 20-25 May, 2024, Torino, Italy}}, \bibfield{editor}{\bibinfo{person}{Nicoletta Calzolari}, \bibinfo{person}{Min{-}Yen Kan}, \bibinfo{person}{V{\'{e}}ronique Hoste}, \bibinfo{person}{Alessandro Lenci}, \bibinfo{person}{Sakriani Sakti}, {and} \bibinfo{person}{Nianwen Xue}} (Eds.). \bibinfo{publisher}{{ELRA} and {ICCL}}, \bibinfo{pages}{16399--16411}.
\newblock
\urldef\tempurl%
\url{https://aclanthology.org/2024.lrec-main.1425}
\showURL{%
\tempurl}


\bibitem[Hannan et~al\mbox{.}(2025)]%
        {10.1007/978-3-031-72664-4_20}
\bibfield{author}{\bibinfo{person}{Tanveer Hannan}, \bibinfo{person}{Md~Mohaiminul Islam}, \bibinfo{person}{Thomas Seidl}, {and} \bibinfo{person}{Gedas Bertasius}.} \bibinfo{year}{2025}\natexlab{}.
\newblock \showarticletitle{RGNet: A Unified Clip Retrieval and Grounding Network for Long Videos}. In \bibinfo{booktitle}{\emph{Computer Vision -- ECCV 2024}}, \bibfield{editor}{\bibinfo{person}{Ale{\v{s}} Leonardis}, \bibinfo{person}{Elisa Ricci}, \bibinfo{person}{Stefan Roth}, \bibinfo{person}{Olga Russakovsky}, \bibinfo{person}{Torsten Sattler}, {and} \bibinfo{person}{G{\"u}l Varol}} (Eds.). \bibinfo{publisher}{Springer Nature Switzerland}, \bibinfo{address}{Cham}, \bibinfo{pages}{352--369}.
\newblock


\bibitem[Hendricks et~al\mbox{.}(2017)]%
        {DBLP:conf/iccv/HendricksWSSDR17}
\bibfield{author}{\bibinfo{person}{Lisa~Anne Hendricks}, \bibinfo{person}{Oliver Wang}, \bibinfo{person}{Eli Shechtman}, \bibinfo{person}{Josef Sivic}, \bibinfo{person}{Trevor Darrell}, {and} \bibinfo{person}{Bryan~C. Russell}.} \bibinfo{year}{2017}\natexlab{}.
\newblock \showarticletitle{Localizing Moments in Video with Natural Language}. In \bibinfo{booktitle}{\emph{{IEEE} International Conference on Computer Vision, {ICCV} 2017, Venice, Italy, October 22-29, 2017}}. \bibinfo{publisher}{{IEEE} Computer Society}, \bibinfo{pages}{5804--5813}.
\newblock
\urldef\tempurl%
\url{https://doi.org/10.1109/ICCV.2017.618}
\showDOI{\tempurl}


\bibitem[Hou et~al\mbox{.}(2021)]%
        {DBLP:conf/mm/HouNC21}
\bibfield{author}{\bibinfo{person}{Zhijian Hou}, \bibinfo{person}{Chong{-}Wah Ngo}, {and} \bibinfo{person}{Wing~Kwong Chan}.} \bibinfo{year}{2021}\natexlab{}.
\newblock \showarticletitle{{CONQUER:} Contextual Query-aware Ranking for Video Corpus Moment Retrieval}. In \bibinfo{booktitle}{\emph{{MM} '21: {ACM} Multimedia Conference, Virtual Event, China, October 20 - 24, 2021}}, \bibfield{editor}{\bibinfo{person}{Heng~Tao Shen}, \bibinfo{person}{Yueting Zhuang}, \bibinfo{person}{John~R. Smith}, \bibinfo{person}{Yang Yang}, \bibinfo{person}{Pablo C{\'{e}}sar}, \bibinfo{person}{Florian Metze}, {and} \bibinfo{person}{Balakrishnan Prabhakaran}} (Eds.). \bibinfo{publisher}{{ACM}}, \bibinfo{pages}{3900--3908}.
\newblock
\urldef\tempurl%
\url{https://doi.org/10.1145/3474085.3475281}
\showDOI{\tempurl}


\bibitem[Hou et~al\mbox{.}(2023)]%
        {DBLP:conf/acl/HouZ0GYCNSD23}
\bibfield{author}{\bibinfo{person}{Zhijian Hou}, \bibinfo{person}{Wanjun Zhong}, \bibinfo{person}{Lei Ji}, \bibinfo{person}{Difei Gao}, \bibinfo{person}{Kun Yan}, \bibinfo{person}{Wing~Kwong Chan}, \bibinfo{person}{Chong{-}Wah Ngo}, \bibinfo{person}{Mike~Zheng Shou}, {and} \bibinfo{person}{Nan Duan}.} \bibinfo{year}{2023}\natexlab{}.
\newblock \showarticletitle{{CONE:} An Efficient COarse-to-fiNE Alignment Framework for Long Video Temporal Grounding}. In \bibinfo{booktitle}{\emph{Proceedings of the 61st Annual Meeting of the Association for Computational Linguistics (Volume 1: Long Papers), {ACL} 2023, Toronto, Canada, July 9-14, 2023}}, \bibfield{editor}{\bibinfo{person}{Anna Rogers}, \bibinfo{person}{Jordan~L. Boyd{-}Graber}, {and} \bibinfo{person}{Naoaki Okazaki}} (Eds.). \bibinfo{publisher}{Association for Computational Linguistics}, \bibinfo{pages}{8013--8028}.
\newblock
\urldef\tempurl%
\url{https://doi.org/10.18653/V1/2023.ACL-LONG.445}
\showDOI{\tempurl}


\bibitem[Indyk and Motwani(1998)]%
        {DBLP:conf/stoc/IndykM98}
\bibfield{author}{\bibinfo{person}{Piotr Indyk} {and} \bibinfo{person}{Rajeev Motwani}.} \bibinfo{year}{1998}\natexlab{}.
\newblock \showarticletitle{Approximate Nearest Neighbors: Towards Removing the Curse of Dimensionality}. In \bibinfo{booktitle}{\emph{Proceedings of the Thirtieth Annual {ACM} Symposium on the Theory of Computing, Dallas, Texas, USA, May 23-26, 1998}}, \bibfield{editor}{\bibinfo{person}{Jeffrey~Scott Vitter}} (Ed.). \bibinfo{publisher}{{ACM}}, \bibinfo{pages}{604--613}.
\newblock
\urldef\tempurl%
\url{https://doi.org/10.1145/276698.276876}
\showDOI{\tempurl}


\bibitem[J{\'{e}}gou et~al\mbox{.}(2011)]%
        {DBLP:journals/pami/JegouDS11}
\bibfield{author}{\bibinfo{person}{Herv{\'{e}} J{\'{e}}gou}, \bibinfo{person}{Matthijs Douze}, {and} \bibinfo{person}{Cordelia Schmid}.} \bibinfo{year}{2011}\natexlab{}.
\newblock \showarticletitle{Product Quantization for Nearest Neighbor Search}.
\newblock \bibinfo{journal}{\emph{{IEEE} Trans. Pattern Anal. Mach. Intell.}} \bibinfo{volume}{33}, \bibinfo{number}{1} (\bibinfo{year}{2011}), \bibinfo{pages}{117--128}.
\newblock
\urldef\tempurl%
\url{https://doi.org/10.1109/TPAMI.2010.57}
\showDOI{\tempurl}


\bibitem[Jia et~al\mbox{.}(2021)]%
        {DBLP:conf/icml/JiaYXCPPLSLD21}
\bibfield{author}{\bibinfo{person}{Chao Jia}, \bibinfo{person}{Yinfei Yang}, \bibinfo{person}{Ye Xia}, \bibinfo{person}{Yi{-}Ting Chen}, \bibinfo{person}{Zarana Parekh}, \bibinfo{person}{Hieu Pham}, \bibinfo{person}{Quoc~V. Le}, \bibinfo{person}{Yun{-}Hsuan Sung}, \bibinfo{person}{Zhen Li}, {and} \bibinfo{person}{Tom Duerig}.} \bibinfo{year}{2021}\natexlab{}.
\newblock \showarticletitle{Scaling Up Visual and Vision-Language Representation Learning With Noisy Text Supervision}. In \bibinfo{booktitle}{\emph{Proceedings of the 38th International Conference on Machine Learning, {ICML} 2021, 18-24 July 2021, Virtual Event}} \emph{(\bibinfo{series}{Proceedings of Machine Learning Research}, Vol.~\bibinfo{volume}{139})}, \bibfield{editor}{\bibinfo{person}{Marina Meila} {and} \bibinfo{person}{Tong Zhang}} (Eds.). \bibinfo{publisher}{{PMLR}}, \bibinfo{pages}{4904--4916}.
\newblock
\urldef\tempurl%
\url{http://proceedings.mlr.press/v139/jia21b.html}
\showURL{%
\tempurl}


\bibitem[Karpukhin et~al\mbox{.}(2020)]%
        {DBLP:conf/emnlp/KarpukhinOMLWEC20}
\bibfield{author}{\bibinfo{person}{Vladimir Karpukhin}, \bibinfo{person}{Barlas Oguz}, \bibinfo{person}{Sewon Min}, \bibinfo{person}{Patrick S.~H. Lewis}, \bibinfo{person}{Ledell Wu}, \bibinfo{person}{Sergey Edunov}, \bibinfo{person}{Danqi Chen}, {and} \bibinfo{person}{Wen{-}tau Yih}.} \bibinfo{year}{2020}\natexlab{}.
\newblock \showarticletitle{Dense Passage Retrieval for Open-Domain Question Answering}. In \bibinfo{booktitle}{\emph{Proceedings of the 2020 Conference on Empirical Methods in Natural Language Processing, {EMNLP} 2020, Online, November 16-20, 2020}}, \bibfield{editor}{\bibinfo{person}{Bonnie Webber}, \bibinfo{person}{Trevor Cohn}, \bibinfo{person}{Yulan He}, {and} \bibinfo{person}{Yang Liu}} (Eds.). \bibinfo{publisher}{Association for Computational Linguistics}, \bibinfo{pages}{6769--6781}.
\newblock
\urldef\tempurl%
\url{https://doi.org/10.18653/V1/2020.EMNLP-MAIN.550}
\showDOI{\tempurl}


\bibitem[Khattab and Zaharia(2020)]%
        {DBLP:conf/sigir/KhattabZ20}
\bibfield{author}{\bibinfo{person}{Omar Khattab} {and} \bibinfo{person}{Matei Zaharia}.} \bibinfo{year}{2020}\natexlab{}.
\newblock \showarticletitle{ColBERT: Efficient and Effective Passage Search via Contextualized Late Interaction over {BERT}}. In \bibinfo{booktitle}{\emph{Proceedings of the 43rd International {ACM} {SIGIR} conference on research and development in Information Retrieval, {SIGIR} 2020, Virtual Event, China, July 25-30, 2020}}, \bibfield{editor}{\bibinfo{person}{Jimmy~X. Huang}, \bibinfo{person}{Yi~Chang}, \bibinfo{person}{Xueqi Cheng}, \bibinfo{person}{Jaap Kamps}, \bibinfo{person}{Vanessa Murdock}, \bibinfo{person}{Ji{-}Rong Wen}, {and} \bibinfo{person}{Yiqun Liu}} (Eds.). \bibinfo{publisher}{{ACM}}, \bibinfo{pages}{39--48}.
\newblock
\urldef\tempurl%
\url{https://doi.org/10.1145/3397271.3401075}
\showDOI{\tempurl}


\bibitem[Krishna et~al\mbox{.}(2017)]%
        {DBLP:conf/iccv/KrishnaHRFN17}
\bibfield{author}{\bibinfo{person}{Ranjay Krishna}, \bibinfo{person}{Kenji Hata}, \bibinfo{person}{Frederic Ren}, \bibinfo{person}{Li Fei{-}Fei}, {and} \bibinfo{person}{Juan~Carlos Niebles}.} \bibinfo{year}{2017}\natexlab{}.
\newblock \showarticletitle{Dense-Captioning Events in Videos}. In \bibinfo{booktitle}{\emph{{IEEE} International Conference on Computer Vision, {ICCV} 2017, Venice, Italy, October 22-29, 2017}}. \bibinfo{publisher}{{IEEE} Computer Society}, \bibinfo{pages}{706--715}.
\newblock
\urldef\tempurl%
\url{https://doi.org/10.1109/ICCV.2017.83}
\showDOI{\tempurl}


\bibitem[Lafferty and Zhai(2001)]%
        {DBLP:conf/sigir/LaffertyZ01}
\bibfield{author}{\bibinfo{person}{John~D. Lafferty} {and} \bibinfo{person}{ChengXiang Zhai}.} \bibinfo{year}{2001}\natexlab{}.
\newblock \showarticletitle{Document Language Models, Query Models, and Risk Minimization for Information Retrieval}. In \bibinfo{booktitle}{\emph{{SIGIR} 2001: Proceedings of the 24th Annual International {ACM} {SIGIR} Conference on Research and Development in Information Retrieval, September 9-13, 2001, New Orleans, Louisiana, {USA}}}, \bibfield{editor}{\bibinfo{person}{W.~Bruce Croft}, \bibinfo{person}{David~J. Harper}, \bibinfo{person}{Donald~H. Kraft}, {and} \bibinfo{person}{Justin Zobel}} (Eds.). \bibinfo{publisher}{{ACM}}, \bibinfo{pages}{111--119}.
\newblock
\urldef\tempurl%
\url{https://doi.org/10.1145/383952.383970}
\showDOI{\tempurl}


\bibitem[Lei et~al\mbox{.}(2020)]%
        {lei2020tvr}
\bibfield{author}{\bibinfo{person}{Jie Lei}, \bibinfo{person}{Licheng Yu}, \bibinfo{person}{Tamara~L Berg}, {and} \bibinfo{person}{Mohit Bansal}.} \bibinfo{year}{2020}\natexlab{}.
\newblock \showarticletitle{Tvr: A large-scale dataset for video-subtitle moment retrieval}. In \bibinfo{booktitle}{\emph{Computer Vision--ECCV 2020: 16th European Conference, Glasgow, UK, August 23--28, 2020, Proceedings, Part XXI 16}}. Springer, \bibinfo{pages}{447--463}.
\newblock


\bibitem[Lewis et~al\mbox{.}(2020)]%
        {DBLP:conf/nips/LewisPPPKGKLYR020}
\bibfield{author}{\bibinfo{person}{Patrick S.~H. Lewis}, \bibinfo{person}{Ethan Perez}, \bibinfo{person}{Aleksandra Piktus}, \bibinfo{person}{Fabio Petroni}, \bibinfo{person}{Vladimir Karpukhin}, \bibinfo{person}{Naman Goyal}, \bibinfo{person}{Heinrich K{\"{u}}ttler}, \bibinfo{person}{Mike Lewis}, \bibinfo{person}{Wen{-}tau Yih}, \bibinfo{person}{Tim Rockt{\"{a}}schel}, \bibinfo{person}{Sebastian Riedel}, {and} \bibinfo{person}{Douwe Kiela}.} \bibinfo{year}{2020}\natexlab{}.
\newblock \showarticletitle{Retrieval-Augmented Generation for Knowledge-Intensive {NLP} Tasks}. In \bibinfo{booktitle}{\emph{Advances in Neural Information Processing Systems 33: Annual Conference on Neural Information Processing Systems 2020, NeurIPS 2020, December 6-12, 2020, virtual}}, \bibfield{editor}{\bibinfo{person}{Hugo Larochelle}, \bibinfo{person}{Marc'Aurelio Ranzato}, \bibinfo{person}{Raia Hadsell}, \bibinfo{person}{Maria{-}Florina Balcan}, {and} \bibinfo{person}{Hsuan{-}Tien Lin}} (Eds.).
\newblock
\urldef\tempurl%
\url{https://proceedings.neurips.cc/paper/2020/hash/6b493230205f780e1bc26945df7481e5-Abstract.html}
\showURL{%
\tempurl}


\bibitem[Li et~al\mbox{.}(2020)]%
        {DBLP:conf/emnlp/LiCCGYL20}
\bibfield{author}{\bibinfo{person}{Linjie Li}, \bibinfo{person}{Yen{-}Chun Chen}, \bibinfo{person}{Yu Cheng}, \bibinfo{person}{Zhe Gan}, \bibinfo{person}{Licheng Yu}, {and} \bibinfo{person}{Jingjing Liu}.} \bibinfo{year}{2020}\natexlab{}.
\newblock \showarticletitle{{HERO:} Hierarchical Encoder for Video+Language Omni-representation Pre-training}. In \bibinfo{booktitle}{\emph{Proceedings of the 2020 Conference on Empirical Methods in Natural Language Processing, {EMNLP} 2020, Online, November 16-20, 2020}}, \bibfield{editor}{\bibinfo{person}{Bonnie Webber}, \bibinfo{person}{Trevor Cohn}, \bibinfo{person}{Yulan He}, {and} \bibinfo{person}{Yang Liu}} (Eds.). \bibinfo{publisher}{Association for Computational Linguistics}, \bibinfo{pages}{2046--2065}.
\newblock
\urldef\tempurl%
\url{https://doi.org/10.18653/V1/2020.EMNLP-MAIN.161}
\showDOI{\tempurl}


\bibitem[Liang et~al\mbox{.}(2024)]%
        {liang2024tvrrankingdatasetrankedvideo}
\bibfield{author}{\bibinfo{person}{Renjie Liang}, \bibinfo{person}{Li Li}, \bibinfo{person}{Chongzhi Zhang}, \bibinfo{person}{Jing Wang}, \bibinfo{person}{Xizhou Zhu}, {and} \bibinfo{person}{Aixin Sun}.} \bibinfo{year}{2024}\natexlab{}.
\newblock \bibinfo{title}{TVR-Ranking: A Dataset for Ranked Video Moment Retrieval with Imprecise Queries}.
\newblock
\newblock
\showeprint[arxiv]{2407.06597}~[cs.AI]
\urldef\tempurl%
\url{https://arxiv.org/abs/2407.06597}
\showURL{%
\tempurl}


\bibitem[Liu(2019)]%
        {liu2019roberta}
\bibfield{author}{\bibinfo{person}{Yinhan Liu}.} \bibinfo{year}{2019}\natexlab{}.
\newblock \showarticletitle{Roberta: A robustly optimized bert pretraining approach}.
\newblock \bibinfo{journal}{\emph{arXiv preprint arXiv:1907.11692}} (\bibinfo{year}{2019}).
\newblock


\bibitem[Liu et~al\mbox{.}(2023)]%
        {DBLP:conf/iclr/LiuXL0023}
\bibfield{author}{\bibinfo{person}{Zhenghao Liu}, \bibinfo{person}{Chenyan Xiong}, \bibinfo{person}{Yuanhuiyi Lv}, \bibinfo{person}{Zhiyuan Liu}, {and} \bibinfo{person}{Ge Yu}.} \bibinfo{year}{2023}\natexlab{}.
\newblock \showarticletitle{Universal Vision-Language Dense Retrieval: Learning {A} Unified Representation Space for Multi-Modal Retrieval}. In \bibinfo{booktitle}{\emph{The Eleventh International Conference on Learning Representations, {ICLR} 2023, Kigali, Rwanda, May 1-5, 2023}}. \bibinfo{publisher}{OpenReview.net}.
\newblock
\urldef\tempurl%
\url{https://openreview.net/forum?id=PQOlkgsBsik}
\showURL{%
\tempurl}


\bibitem[Loshchilov and Hutter(2019)]%
        {DBLP:conf/iclr/LoshchilovH19}
\bibfield{author}{\bibinfo{person}{Ilya Loshchilov} {and} \bibinfo{person}{Frank Hutter}.} \bibinfo{year}{2019}\natexlab{}.
\newblock \showarticletitle{Decoupled Weight Decay Regularization}. In \bibinfo{booktitle}{\emph{7th International Conference on Learning Representations, {ICLR} 2019, New Orleans, LA, USA, May 6-9, 2019}}. \bibinfo{publisher}{OpenReview.net}.
\newblock
\urldef\tempurl%
\url{https://openreview.net/forum?id=Bkg6RiCqY7}
\showURL{%
\tempurl}


\bibitem[Lou et~al\mbox{.}(2022)]%
        {DBLP:conf/icassp/LouXWY22}
\bibfield{author}{\bibinfo{person}{Siyu Lou}, \bibinfo{person}{Xuenan Xu}, \bibinfo{person}{Mengyue Wu}, {and} \bibinfo{person}{Kai Yu}.} \bibinfo{year}{2022}\natexlab{}.
\newblock \showarticletitle{Audio-Text Retrieval in Context}. In \bibinfo{booktitle}{\emph{{IEEE} International Conference on Acoustics, Speech and Signal Processing, {ICASSP} 2022, Virtual and Singapore, 23-27 May 2022}}. \bibinfo{publisher}{{IEEE}}, \bibinfo{pages}{4793--4797}.
\newblock
\urldef\tempurl%
\url{https://doi.org/10.1109/ICASSP43922.2022.9746786}
\showDOI{\tempurl}


\bibitem[Luo et~al\mbox{.}(2022)]%
        {DBLP:journals/ijon/LuoJZCLDL22}
\bibfield{author}{\bibinfo{person}{Huaishao Luo}, \bibinfo{person}{Lei Ji}, \bibinfo{person}{Ming Zhong}, \bibinfo{person}{Yang Chen}, \bibinfo{person}{Wen Lei}, \bibinfo{person}{Nan Duan}, {and} \bibinfo{person}{Tianrui Li}.} \bibinfo{year}{2022}\natexlab{}.
\newblock \showarticletitle{CLIP4Clip: An empirical study of {CLIP} for end to end video clip retrieval and captioning}.
\newblock \bibinfo{journal}{\emph{Neurocomputing}}  \bibinfo{volume}{508} (\bibinfo{year}{2022}), \bibinfo{pages}{293--304}.
\newblock
\urldef\tempurl%
\url{https://doi.org/10.1016/J.NEUCOM.2022.07.028}
\showDOI{\tempurl}


\bibitem[Macdonald et~al\mbox{.}(2013)]%
        {DBLP:journals/ir/MacdonaldSO13}
\bibfield{author}{\bibinfo{person}{Craig Macdonald}, \bibinfo{person}{Rodrygo L.~T. Santos}, {and} \bibinfo{person}{Iadh Ounis}.} \bibinfo{year}{2013}\natexlab{}.
\newblock \showarticletitle{The whens and hows of learning to rank for web search}.
\newblock \bibinfo{journal}{\emph{Inf. Retr.}} \bibinfo{volume}{16}, \bibinfo{number}{5} (\bibinfo{year}{2013}), \bibinfo{pages}{584--628}.
\newblock
\urldef\tempurl%
\url{https://doi.org/10.1007/S10791-012-9209-9}
\showDOI{\tempurl}


\bibitem[Malkov and Yashunin(2020)]%
        {DBLP:journals/pami/MalkovY20}
\bibfield{author}{\bibinfo{person}{Yury~A. Malkov} {and} \bibinfo{person}{Dmitry~A. Yashunin}.} \bibinfo{year}{2020}\natexlab{}.
\newblock \showarticletitle{Efficient and Robust Approximate Nearest Neighbor Search Using Hierarchical Navigable Small World Graphs}.
\newblock \bibinfo{journal}{\emph{{IEEE} Trans. Pattern Anal. Mach. Intell.}} \bibinfo{volume}{42}, \bibinfo{number}{4} (\bibinfo{year}{2020}), \bibinfo{pages}{824--836}.
\newblock
\urldef\tempurl%
\url{https://doi.org/10.1109/TPAMI.2018.2889473}
\showDOI{\tempurl}


\bibitem[Miech et~al\mbox{.}(2020)]%
        {DBLP:conf/cvpr/MiechASLSZ20}
\bibfield{author}{\bibinfo{person}{Antoine Miech}, \bibinfo{person}{Jean{-}Baptiste Alayrac}, \bibinfo{person}{Lucas Smaira}, \bibinfo{person}{Ivan Laptev}, \bibinfo{person}{Josef Sivic}, {and} \bibinfo{person}{Andrew Zisserman}.} \bibinfo{year}{2020}\natexlab{}.
\newblock \showarticletitle{End-to-End Learning of Visual Representations From Uncurated Instructional Videos}. In \bibinfo{booktitle}{\emph{2020 {IEEE/CVF} Conference on Computer Vision and Pattern Recognition, {CVPR} 2020, Seattle, WA, USA, June 13-19, 2020}}. \bibinfo{publisher}{Computer Vision Foundation / {IEEE}}, \bibinfo{pages}{9876--9886}.
\newblock
\urldef\tempurl%
\url{https://doi.org/10.1109/CVPR42600.2020.00990}
\showDOI{\tempurl}


\bibitem[Mithun et~al\mbox{.}(2018)]%
        {DBLP:conf/mir/MithunLMR18}
\bibfield{author}{\bibinfo{person}{Niluthpol~Chowdhury Mithun}, \bibinfo{person}{Juncheng Li}, \bibinfo{person}{Florian Metze}, {and} \bibinfo{person}{Amit~K. Roy{-}Chowdhury}.} \bibinfo{year}{2018}\natexlab{}.
\newblock \showarticletitle{Learning Joint Embedding with Multimodal Cues for Cross-Modal Video-Text Retrieval}. In \bibinfo{booktitle}{\emph{Proceedings of the 2018 {ACM} on International Conference on Multimedia Retrieval, {ICMR} 2018, Yokohama, Japan, June 11-14, 2018}}, \bibfield{editor}{\bibinfo{person}{Kiyoharu Aizawa}, \bibinfo{person}{Michael~S. Lew}, {and} \bibinfo{person}{Shin'ichi Satoh}} (Eds.). \bibinfo{publisher}{{ACM}}, \bibinfo{pages}{19--27}.
\newblock
\urldef\tempurl%
\url{https://doi.org/10.1145/3206025.3206064}
\showDOI{\tempurl}


\bibitem[Neyshabur and Srebro(2015)]%
        {DBLP:conf/icml/NeyshaburS15}
\bibfield{author}{\bibinfo{person}{Behnam Neyshabur} {and} \bibinfo{person}{Nathan Srebro}.} \bibinfo{year}{2015}\natexlab{}.
\newblock \showarticletitle{On Symmetric and Asymmetric LSHs for Inner Product Search}. In \bibinfo{booktitle}{\emph{Proceedings of the 32nd International Conference on Machine Learning, {ICML} 2015, Lille, France, 6-11 July 2015}} \emph{(\bibinfo{series}{{JMLR} Workshop and Conference Proceedings}, Vol.~\bibinfo{volume}{37})}, \bibfield{editor}{\bibinfo{person}{Francis~R. Bach} {and} \bibinfo{person}{David~M. Blei}} (Eds.). \bibinfo{publisher}{JMLR.org}, \bibinfo{pages}{1926--1934}.
\newblock
\urldef\tempurl%
\url{http://proceedings.mlr.press/v37/neyshabur15.html}
\showURL{%
\tempurl}


\bibitem[Pan et~al\mbox{.}(2023)]%
        {DBLP:conf/iccv/PanHGLSPZ23}
\bibfield{author}{\bibinfo{person}{Yulin Pan}, \bibinfo{person}{Xiangteng He}, \bibinfo{person}{Biao Gong}, \bibinfo{person}{Yiliang Lv}, \bibinfo{person}{Yujun Shen}, \bibinfo{person}{Yuxin Peng}, {and} \bibinfo{person}{Deli Zhao}.} \bibinfo{year}{2023}\natexlab{}.
\newblock \showarticletitle{Scanning Only Once: An End-to-end Framework for Fast Temporal Grounding in Long Videos}. In \bibinfo{booktitle}{\emph{{IEEE/CVF} International Conference on Computer Vision, {ICCV} 2023, Paris, France, October 1-6, 2023}}. \bibinfo{publisher}{{IEEE}}, \bibinfo{pages}{13721--13731}.
\newblock
\urldef\tempurl%
\url{https://doi.org/10.1109/ICCV51070.2023.01266}
\showDOI{\tempurl}


\bibitem[Qu et~al\mbox{.}(2020)]%
        {DBLP:conf/mm/QuTZ0DZX20}
\bibfield{author}{\bibinfo{person}{Xiaoye Qu}, \bibinfo{person}{Pengwei Tang}, \bibinfo{person}{Zhikang Zou}, \bibinfo{person}{Yu Cheng}, \bibinfo{person}{Jianfeng Dong}, \bibinfo{person}{Pan Zhou}, {and} \bibinfo{person}{Zichuan Xu}.} \bibinfo{year}{2020}\natexlab{}.
\newblock \showarticletitle{Fine-grained Iterative Attention Network for Temporal Language Localization in Videos}. In \bibinfo{booktitle}{\emph{{MM} '20: The 28th {ACM} International Conference on Multimedia, Virtual Event / Seattle, WA, USA, October 12-16, 2020}}, \bibfield{editor}{\bibinfo{person}{Chang~Wen Chen}, \bibinfo{person}{Rita Cucchiara}, \bibinfo{person}{Xian{-}Sheng Hua}, \bibinfo{person}{Guo{-}Jun Qi}, \bibinfo{person}{Elisa Ricci}, \bibinfo{person}{Zhengyou Zhang}, {and} \bibinfo{person}{Roger Zimmermann}} (Eds.). \bibinfo{publisher}{{ACM}}, \bibinfo{pages}{4280--4288}.
\newblock
\urldef\tempurl%
\url{https://doi.org/10.1145/3394171.3414053}
\showDOI{\tempurl}


\bibitem[Radford et~al\mbox{.}(2021)]%
        {radford2021learning}
\bibfield{author}{\bibinfo{person}{Alec Radford}, \bibinfo{person}{Jong~Wook Kim}, \bibinfo{person}{Chris Hallacy}, \bibinfo{person}{Aditya Ramesh}, \bibinfo{person}{Gabriel Goh}, \bibinfo{person}{Sandhini Agarwal}, \bibinfo{person}{Girish Sastry}, \bibinfo{person}{Amanda Askell}, \bibinfo{person}{Pamela Mishkin}, \bibinfo{person}{Jack Clark}, {et~al\mbox{.}}} \bibinfo{year}{2021}\natexlab{}.
\newblock \showarticletitle{Learning transferable visual models from natural language supervision}. In \bibinfo{booktitle}{\emph{International conference on machine learning}}. PMLR, \bibinfo{pages}{8748--8763}.
\newblock


\bibitem[Regneri et~al\mbox{.}(2013)]%
        {DBLP:journals/tacl/RegneriRWTSP13}
\bibfield{author}{\bibinfo{person}{Michaela Regneri}, \bibinfo{person}{Marcus Rohrbach}, \bibinfo{person}{Dominikus Wetzel}, \bibinfo{person}{Stefan Thater}, \bibinfo{person}{Bernt Schiele}, {and} \bibinfo{person}{Manfred Pinkal}.} \bibinfo{year}{2013}\natexlab{}.
\newblock \showarticletitle{Grounding Action Descriptions in Videos}.
\newblock \bibinfo{journal}{\emph{Trans. Assoc. Comput. Linguistics}}  \bibinfo{volume}{1} (\bibinfo{year}{2013}), \bibinfo{pages}{25--36}.
\newblock
\urldef\tempurl%
\url{https://doi.org/10.1162/TACL\_A\_00207}
\showDOI{\tempurl}


\bibitem[Robertson and Walker(1997)]%
        {DBLP:conf/sigir/RobertsonW97}
\bibfield{author}{\bibinfo{person}{Stephen~E. Robertson} {and} \bibinfo{person}{Steve Walker}.} \bibinfo{year}{1997}\natexlab{}.
\newblock \showarticletitle{On Relevance Weights with Little Relevance Information}. In \bibinfo{booktitle}{\emph{{SIGIR} '97: Proceedings of the 20th Annual International {ACM} {SIGIR} Conference on Research and Development in Information Retrieval, July 27-31, 1997, Philadelphia, PA, {USA}}}, \bibfield{editor}{\bibinfo{person}{Nicholas~J. Belkin}, \bibinfo{person}{Arcot~Desai Narasimhalu}, \bibinfo{person}{Peter Willett}, \bibinfo{person}{William~R. Hersh}, \bibinfo{person}{Fazli Can}, {and} \bibinfo{person}{Ellen~M. Voorhees}} (Eds.). \bibinfo{publisher}{{ACM}}, \bibinfo{pages}{16--24}.
\newblock
\urldef\tempurl%
\url{https://doi.org/10.1145/258525.258529}
\showDOI{\tempurl}


\bibitem[Robertson and Zaragoza(2009)]%
        {DBLP:journals/ftir/RobertsonZ09}
\bibfield{author}{\bibinfo{person}{Stephen~E. Robertson} {and} \bibinfo{person}{Hugo Zaragoza}.} \bibinfo{year}{2009}\natexlab{}.
\newblock \showarticletitle{The Probabilistic Relevance Framework: {BM25} and Beyond}.
\newblock \bibinfo{journal}{\emph{Found. Trends Inf. Retr.}} \bibinfo{volume}{3}, \bibinfo{number}{4} (\bibinfo{year}{2009}), \bibinfo{pages}{333--389}.
\newblock
\urldef\tempurl%
\url{https://doi.org/10.1561/1500000019}
\showDOI{\tempurl}


\bibitem[Sigurdsson et~al\mbox{.}(2016)]%
        {DBLP:conf/eccv/SigurdssonVWFLG16}
\bibfield{author}{\bibinfo{person}{Gunnar~A. Sigurdsson}, \bibinfo{person}{G{\"{u}}l Varol}, \bibinfo{person}{Xiaolong Wang}, \bibinfo{person}{Ali Farhadi}, \bibinfo{person}{Ivan Laptev}, {and} \bibinfo{person}{Abhinav Gupta}.} \bibinfo{year}{2016}\natexlab{}.
\newblock \showarticletitle{Hollywood in Homes: Crowdsourcing Data Collection for Activity Understanding}. In \bibinfo{booktitle}{\emph{Computer Vision - {ECCV} 2016 - 14th European Conference, Amsterdam, The Netherlands, October 11-14, 2016, Proceedings, Part {I}}} \emph{(\bibinfo{series}{Lecture Notes in Computer Science}, Vol.~\bibinfo{volume}{9905})}, \bibfield{editor}{\bibinfo{person}{Bastian Leibe}, \bibinfo{person}{Jiri Matas}, \bibinfo{person}{Nicu Sebe}, {and} \bibinfo{person}{Max Welling}} (Eds.). \bibinfo{publisher}{Springer}, \bibinfo{pages}{510--526}.
\newblock
\urldef\tempurl%
\url{https://doi.org/10.1007/978-3-319-46448-0\_31}
\showDOI{\tempurl}


\bibitem[Soldan et~al\mbox{.}(2022)]%
        {DBLP:conf/cvpr/SoldanPAH0GG22}
\bibfield{author}{\bibinfo{person}{Mattia Soldan}, \bibinfo{person}{Alejandro Pardo}, \bibinfo{person}{Juan~Le{\'{o}}n Alc{\'{a}}zar}, \bibinfo{person}{Fabian~Caba Heilbron}, \bibinfo{person}{Chen Zhao}, \bibinfo{person}{Silvio Giancola}, {and} \bibinfo{person}{Bernard Ghanem}.} \bibinfo{year}{2022}\natexlab{}.
\newblock \showarticletitle{{MAD:} {A} Scalable Dataset for Language Grounding in Videos from Movie Audio Descriptions}. In \bibinfo{booktitle}{\emph{{IEEE/CVF} Conference on Computer Vision and Pattern Recognition, {CVPR} 2022, New Orleans, LA, USA, June 18-24, 2022}}. \bibinfo{publisher}{{IEEE}}, \bibinfo{pages}{5016--5025}.
\newblock
\urldef\tempurl%
\url{https://doi.org/10.1109/CVPR52688.2022.00497}
\showDOI{\tempurl}


\bibitem[Tonellotto(2022)]%
        {DBLP:journals/corr/abs-2207-13443}
\bibfield{author}{\bibinfo{person}{Nicola Tonellotto}.} \bibinfo{year}{2022}\natexlab{}.
\newblock \showarticletitle{Lecture Notes on Neural Information Retrieval}.
\newblock \bibinfo{journal}{\emph{CoRR}}  \bibinfo{volume}{abs/2207.13443} (\bibinfo{year}{2022}).
\newblock
\urldef\tempurl%
\url{https://doi.org/10.48550/ARXIV.2207.13443}
\showDOI{\tempurl}
\showeprint[arXiv]{2207.13443}


\bibitem[Wang et~al\mbox{.}(2021)]%
        {DBLP:conf/cvpr/WangZL0L21}
\bibfield{author}{\bibinfo{person}{Hao Wang}, \bibinfo{person}{Zheng{-}Jun Zha}, \bibinfo{person}{Liang Li}, \bibinfo{person}{Dong Liu}, {and} \bibinfo{person}{Jiebo Luo}.} \bibinfo{year}{2021}\natexlab{}.
\newblock \showarticletitle{Structured Multi-Level Interaction Network for Video Moment Localization via Language Query}. In \bibinfo{booktitle}{\emph{{IEEE} Conference on Computer Vision and Pattern Recognition, {CVPR} 2021, virtual, June 19-25, 2021}}. \bibinfo{publisher}{Computer Vision Foundation / {IEEE}}, \bibinfo{pages}{7026--7035}.
\newblock
\urldef\tempurl%
\url{https://doi.org/10.1109/CVPR46437.2021.00695}
\showDOI{\tempurl}


\bibitem[Wu et~al\mbox{.}(2023)]%
        {DBLP:conf/icassp/WuCZHBD23}
\bibfield{author}{\bibinfo{person}{Yusong Wu}, \bibinfo{person}{Ke Chen}, \bibinfo{person}{Tianyu Zhang}, \bibinfo{person}{Yuchen Hui}, \bibinfo{person}{Taylor Berg{-}Kirkpatrick}, {and} \bibinfo{person}{Shlomo Dubnov}.} \bibinfo{year}{2023}\natexlab{}.
\newblock \showarticletitle{Large-Scale Contrastive Language-Audio Pretraining with Feature Fusion and Keyword-to-Caption Augmentation}. In \bibinfo{booktitle}{\emph{{IEEE} International Conference on Acoustics, Speech and Signal Processing {ICASSP} 2023, Rhodes Island, Greece, June 4-10, 2023}}. \bibinfo{publisher}{{IEEE}}, \bibinfo{pages}{1--5}.
\newblock
\urldef\tempurl%
\url{https://doi.org/10.1109/ICASSP49357.2023.10095969}
\showDOI{\tempurl}


\bibitem[Xiao et~al\mbox{.}(2021)]%
        {DBLP:conf/aaai/XiaoCZJSYX21}
\bibfield{author}{\bibinfo{person}{Shaoning Xiao}, \bibinfo{person}{Long Chen}, \bibinfo{person}{Songyang Zhang}, \bibinfo{person}{Wei Ji}, \bibinfo{person}{Jian Shao}, \bibinfo{person}{Lu Ye}, {and} \bibinfo{person}{Jun Xiao}.} \bibinfo{year}{2021}\natexlab{}.
\newblock \showarticletitle{Boundary Proposal Network for Two-stage Natural Language Video Localization}. In \bibinfo{booktitle}{\emph{Thirty-Fifth {AAAI} Conference on Artificial Intelligence, {AAAI} 2021, Thirty-Third Conference on Innovative Applications of Artificial Intelligence, {IAAI} 2021, The Eleventh Symposium on Educational Advances in Artificial Intelligence, {EAAI} 2021, Virtual Event, February 2-9, 2021}}. \bibinfo{publisher}{{AAAI} Press}, \bibinfo{pages}{2986--2994}.
\newblock
\urldef\tempurl%
\url{https://doi.org/10.1609/AAAI.V35I4.16406}
\showDOI{\tempurl}


\bibitem[Xiong et~al\mbox{.}(2021)]%
        {DBLP:conf/iclr/XiongXLTLBAO21}
\bibfield{author}{\bibinfo{person}{Lee Xiong}, \bibinfo{person}{Chenyan Xiong}, \bibinfo{person}{Ye Li}, \bibinfo{person}{Kwok{-}Fung Tang}, \bibinfo{person}{Jialin Liu}, \bibinfo{person}{Paul~N. Bennett}, \bibinfo{person}{Junaid Ahmed}, {and} \bibinfo{person}{Arnold Overwijk}.} \bibinfo{year}{2021}\natexlab{}.
\newblock \showarticletitle{Approximate Nearest Neighbor Negative Contrastive Learning for Dense Text Retrieval}. In \bibinfo{booktitle}{\emph{9th International Conference on Learning Representations, {ICLR} 2021, Virtual Event, Austria, May 3-7, 2021}}. \bibinfo{publisher}{OpenReview.net}.
\newblock
\urldef\tempurl%
\url{https://openreview.net/forum?id=zeFrfgyZln}
\showURL{%
\tempurl}


\bibitem[Xu et~al\mbox{.}(2021)]%
        {DBLP:conf/emnlp/XuG0OAMZF21}
\bibfield{author}{\bibinfo{person}{Hu Xu}, \bibinfo{person}{Gargi Ghosh}, \bibinfo{person}{Po{-}Yao Huang}, \bibinfo{person}{Dmytro Okhonko}, \bibinfo{person}{Armen Aghajanyan}, \bibinfo{person}{Florian Metze}, \bibinfo{person}{Luke Zettlemoyer}, {and} \bibinfo{person}{Christoph Feichtenhofer}.} \bibinfo{year}{2021}\natexlab{}.
\newblock \showarticletitle{VideoCLIP: Contrastive Pre-training for Zero-shot Video-Text Understanding}. In \bibinfo{booktitle}{\emph{Proceedings of the 2021 Conference on Empirical Methods in Natural Language Processing, {EMNLP} 2021, Virtual Event / Punta Cana, Dominican Republic, 7-11 November, 2021}}, \bibfield{editor}{\bibinfo{person}{Marie{-}Francine Moens}, \bibinfo{person}{Xuanjing Huang}, \bibinfo{person}{Lucia Specia}, {and} \bibinfo{person}{Scott~Wen{-}tau Yih}} (Eds.). \bibinfo{publisher}{Association for Computational Linguistics}, \bibinfo{pages}{6787--6800}.
\newblock
\urldef\tempurl%
\url{https://doi.org/10.18653/V1/2021.EMNLP-MAIN.544}
\showDOI{\tempurl}


\bibitem[Xu et~al\mbox{.}(2019)]%
        {DBLP:conf/aaai/Xu0PSSS19}
\bibfield{author}{\bibinfo{person}{Huijuan Xu}, \bibinfo{person}{Kun He}, \bibinfo{person}{Bryan~A. Plummer}, \bibinfo{person}{Leonid Sigal}, \bibinfo{person}{Stan Sclaroff}, {and} \bibinfo{person}{Kate Saenko}.} \bibinfo{year}{2019}\natexlab{}.
\newblock \showarticletitle{Multilevel Language and Vision Integration for Text-to-Clip Retrieval}. In \bibinfo{booktitle}{\emph{The Thirty-Third {AAAI} Conference on Artificial Intelligence, {AAAI} 2019, The Thirty-First Innovative Applications of Artificial Intelligence Conference, {IAAI} 2019, The Ninth {AAAI} Symposium on Educational Advances in Artificial Intelligence, {EAAI} 2019, Honolulu, Hawaii, USA, January 27 - February 1, 2019}}. \bibinfo{publisher}{{AAAI} Press}, \bibinfo{pages}{9062--9069}.
\newblock
\urldef\tempurl%
\url{https://doi.org/10.1609/AAAI.V33I01.33019062}
\showDOI{\tempurl}


\bibitem[Yuan et~al\mbox{.}(2024)]%
        {DBLP:conf/cvpr/YuanZLLCJJ24}
\bibfield{author}{\bibinfo{person}{Tongtong Yuan}, \bibinfo{person}{Xuange Zhang}, \bibinfo{person}{Kun Liu}, \bibinfo{person}{Bo Liu}, \bibinfo{person}{Chen Chen}, \bibinfo{person}{Jian Jin}, {and} \bibinfo{person}{Zhenzhen Jiao}.} \bibinfo{year}{2024}\natexlab{}.
\newblock \showarticletitle{Towards Surveillance Video-and-Language Understanding: New Dataset, Baselines, and Challenges}. In \bibinfo{booktitle}{\emph{{IEEE/CVF} Conference on Computer Vision and Pattern Recognition, {CVPR} 2024, Seattle, WA, USA, June 16-22, 2024}}. \bibinfo{publisher}{{IEEE}}, \bibinfo{pages}{22052--22061}.
\newblock
\urldef\tempurl%
\url{https://doi.org/10.1109/CVPR52733.2024.02082}
\showDOI{\tempurl}


\bibitem[Zhan et~al\mbox{.}(2021)]%
        {DBLP:conf/sigir/ZhanM0G0M21}
\bibfield{author}{\bibinfo{person}{Jingtao Zhan}, \bibinfo{person}{Jiaxin Mao}, \bibinfo{person}{Yiqun Liu}, \bibinfo{person}{Jiafeng Guo}, \bibinfo{person}{Min Zhang}, {and} \bibinfo{person}{Shaoping Ma}.} \bibinfo{year}{2021}\natexlab{}.
\newblock \showarticletitle{Optimizing Dense Retrieval Model Training with Hard Negatives}. In \bibinfo{booktitle}{\emph{{SIGIR} '21: The 44th International {ACM} {SIGIR} Conference on Research and Development in Information Retrieval, Virtual Event, Canada, July 11-15, 2021}}, \bibfield{editor}{\bibinfo{person}{Fernando Diaz}, \bibinfo{person}{Chirag Shah}, \bibinfo{person}{Torsten Suel}, \bibinfo{person}{Pablo Castells}, \bibinfo{person}{Rosie Jones}, {and} \bibinfo{person}{Tetsuya Sakai}} (Eds.). \bibinfo{publisher}{{ACM}}, \bibinfo{pages}{1503--1512}.
\newblock
\urldef\tempurl%
\url{https://doi.org/10.1145/3404835.3462880}
\showDOI{\tempurl}


\bibitem[Zhang et~al\mbox{.}(2021a)]%
        {DBLP:conf/sigir/0048SJNZZG21}
\bibfield{author}{\bibinfo{person}{Hao Zhang}, \bibinfo{person}{Aixin Sun}, \bibinfo{person}{Wei Jing}, \bibinfo{person}{Guoshun Nan}, \bibinfo{person}{Liangli Zhen}, \bibinfo{person}{Joey~Tianyi Zhou}, {and} \bibinfo{person}{Rick Siow~Mong Goh}.} \bibinfo{year}{2021}\natexlab{a}.
\newblock \showarticletitle{Video Corpus Moment Retrieval with Contrastive Learning}. In \bibinfo{booktitle}{\emph{{SIGIR} '21: The 44th International {ACM} {SIGIR} Conference on Research and Development in Information Retrieval, Virtual Event, Canada, July 11-15, 2021}}, \bibfield{editor}{\bibinfo{person}{Fernando Diaz}, \bibinfo{person}{Chirag Shah}, \bibinfo{person}{Torsten Suel}, \bibinfo{person}{Pablo Castells}, \bibinfo{person}{Rosie Jones}, {and} \bibinfo{person}{Tetsuya Sakai}} (Eds.). \bibinfo{publisher}{{ACM}}, \bibinfo{pages}{685--695}.
\newblock
\urldef\tempurl%
\url{https://doi.org/10.1145/3404835.3462874}
\showDOI{\tempurl}


\bibitem[Zhang et~al\mbox{.}(2021b)]%
        {DBLP:conf/acl/ZhangSJZZG21}
\bibfield{author}{\bibinfo{person}{Hao Zhang}, \bibinfo{person}{Aixin Sun}, \bibinfo{person}{Wei Jing}, \bibinfo{person}{Liangli Zhen}, \bibinfo{person}{Joey~Tianyi Zhou}, {and} \bibinfo{person}{Rick Siow~Mong Goh}.} \bibinfo{year}{2021}\natexlab{b}.
\newblock \showarticletitle{Parallel Attention Network with Sequence Matching for Video Grounding}. In \bibinfo{booktitle}{\emph{Findings of the Association for Computational Linguistics: {ACL/IJCNLP} 2021, Online Event, August 1-6, 2021}} \emph{(\bibinfo{series}{Findings of {ACL}}, Vol.~\bibinfo{volume}{{ACL/IJCNLP} 2021})}, \bibfield{editor}{\bibinfo{person}{Chengqing Zong}, \bibinfo{person}{Fei Xia}, \bibinfo{person}{Wenjie Li}, {and} \bibinfo{person}{Roberto Navigli}} (Eds.). \bibinfo{publisher}{Association for Computational Linguistics}, \bibinfo{pages}{776--790}.
\newblock
\urldef\tempurl%
\url{https://doi.org/10.18653/V1/2021.FINDINGS-ACL.69}
\showDOI{\tempurl}


\bibitem[Zhang et~al\mbox{.}(2020b)]%
        {DBLP:conf/acl/ZhangSJZ20}
\bibfield{author}{\bibinfo{person}{Hao Zhang}, \bibinfo{person}{Aixin Sun}, \bibinfo{person}{Wei Jing}, {and} \bibinfo{person}{Joey~Tianyi Zhou}.} \bibinfo{year}{2020}\natexlab{b}.
\newblock \showarticletitle{Span-based Localizing Network for Natural Language Video Localization}. In \bibinfo{booktitle}{\emph{Proceedings of the 58th Annual Meeting of the Association for Computational Linguistics, {ACL} 2020, Online, July 5-10, 2020}}, \bibfield{editor}{\bibinfo{person}{Dan Jurafsky}, \bibinfo{person}{Joyce Chai}, \bibinfo{person}{Natalie Schluter}, {and} \bibinfo{person}{Joel~R. Tetreault}} (Eds.). \bibinfo{publisher}{Association for Computational Linguistics}, \bibinfo{pages}{6543--6554}.
\newblock
\urldef\tempurl%
\url{https://doi.org/10.18653/V1/2020.ACL-MAIN.585}
\showDOI{\tempurl}


\bibitem[Zhang et~al\mbox{.}(2023)]%
        {DBLP:conf/iccv/ZhangGPCHLYL23}
\bibfield{author}{\bibinfo{person}{Mingyuan Zhang}, \bibinfo{person}{Xinying Guo}, \bibinfo{person}{Liang Pan}, \bibinfo{person}{Zhongang Cai}, \bibinfo{person}{Fangzhou Hong}, \bibinfo{person}{Huirong Li}, \bibinfo{person}{Lei Yang}, {and} \bibinfo{person}{Ziwei Liu}.} \bibinfo{year}{2023}\natexlab{}.
\newblock \showarticletitle{ReMoDiffuse: Retrieval-Augmented Motion Diffusion Model}. In \bibinfo{booktitle}{\emph{{IEEE/CVF} International Conference on Computer Vision, {ICCV} 2023, Paris, France, October 1-6, 2023}}. \bibinfo{publisher}{{IEEE}}, \bibinfo{pages}{364--373}.
\newblock
\urldef\tempurl%
\url{https://doi.org/10.1109/ICCV51070.2023.00040}
\showDOI{\tempurl}


\bibitem[Zhang et~al\mbox{.}(2020a)]%
        {DBLP:conf/aaai/ZhangPFL20}
\bibfield{author}{\bibinfo{person}{Songyang Zhang}, \bibinfo{person}{Houwen Peng}, \bibinfo{person}{Jianlong Fu}, {and} \bibinfo{person}{Jiebo Luo}.} \bibinfo{year}{2020}\natexlab{a}.
\newblock \showarticletitle{Learning 2D Temporal Adjacent Networks for Moment Localization with Natural Language}. In \bibinfo{booktitle}{\emph{The Thirty-Fourth {AAAI} Conference on Artificial Intelligence, {AAAI} 2020, The Thirty-Second Innovative Applications of Artificial Intelligence Conference, {IAAI} 2020, The Tenth {AAAI} Symposium on Educational Advances in Artificial Intelligence, {EAAI} 2020, New York, NY, USA, February 7-12, 2020}}. \bibinfo{publisher}{{AAAI} Press}, \bibinfo{pages}{12870--12877}.
\newblock
\urldef\tempurl%
\url{https://doi.org/10.1609/AAAI.V34I07.6984}
\showDOI{\tempurl}


\end{thebibliography}

\appendix

%======================
\begin{figure*}
         \includegraphics[width=\linewidth]{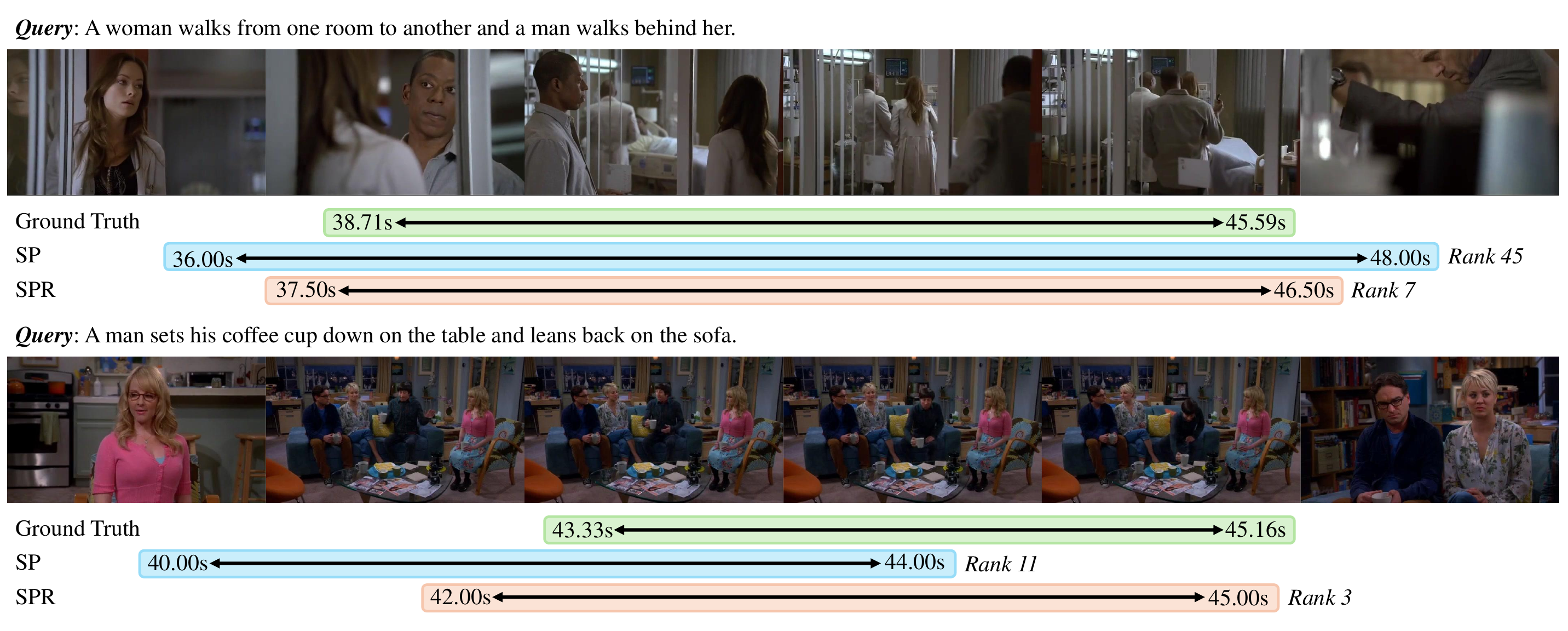}
    \caption{\textbf{Visualization of the ground truth moment and the results from SP and SPR} for two example queries from the TVR-Ranking dataset. SP generates coarse proposals by aggregating retrieved relevant segments, while SPR refines and re-ranks these proposals. Although SP retrieves highly relevant moments, its timestamps are constrained by the pre-defined segment length (\eg 4 seconds in our setting). Hence, all proposals from SP have a length that is a multiple of 4 seconds. In contrast, SPR identifies more precise timestamps and ranks relevant moments more effectively.}
   
    \label{fig:qual_result}
\end{figure*}
%======================

%===============================
\section{Qualitative Results}
%===============================
As a case study, in \cref{fig:qual_result},  we present two queries, each paired with a sampled video moment from the TVR-Ranking dataset~\cite{liang2024tvrrankingdatasetrankedvideo}. In both examples, the sampled moments are highly relevant to the queries, with relevance scores of 4 and 3, respectively.

We plot the ground truth moment as a reference, along with the start/end timestamps and rankings obtained from SP and SPR, respectively. The SP results are coarse proposals aggregated from the retrieved relevant segments; hence, their length is constrained by the pre-defined segment length, which is a multiple of 4 seconds in our setting. The SPR results, specifically from the best configuration, SPR$_{\text{ReLo}}$-L, represent the refined and ranked outputs.

As visualized in \cref{fig:qual_result}, we observe the following: (i) The SP model retrieves highly relevant moments, but it lacks temporal precision due to the minimum time scale of the segment $\tau_{S}$. Consequently, its predicted timestamps do not sufficiently overlap with the ground truth. In the first example, it includes excessive context at the beginning and end, while in the second example, it misses parts of the ground truth. (ii) The SPR model manages to predict more precise timestamps and rank these highly relevant moments with better ranking positions. 

%======================
\begin{table*}[t!]
\centering
\caption{\textbf{Ablation study on the impact of retrieving a different number of segments}. The results indicate that performance peaks at 200 segments across most metrics. Therefore, we set the default number of segments for retrieval to 200 in all experiments to achieve optimal performance and efficiency.}
\resizebox{1.0\linewidth}{!}{
\begin{tabular}{lcccccccccc}
\toprule
\multirow{2}{*}{\textbf{Seg. R@}} & \multicolumn{3}{c}{$\bm{\textbf{NDCG}@10}$} & \multicolumn{3}{c}{$\bm{\textbf{NDCG}@20}$} & \multicolumn{3}{c}{$\bm{\textbf{NDCG}@40}$} \\
\cmidrule(lr){2-4} \cmidrule(lr){5-7} \cmidrule(lr){8-10}
 & $\bm{\textbf{IoU}\geq 0.3}$ & $\bm{\textbf{IoU}\geq 0.5}$ & $\bm{\textbf{IoU}\geq 0.7}$ & $\bm{\textbf{IoU}\geq 0.3}$ & $\bm{\textbf{IoU}\geq 0.5}$ & $\bm{\textbf{IoU}\geq 0.7}$ & $\bm{\textbf{IoU}\geq 0.3}$ & $\bm{\textbf{IoU}\geq 0.5}$ & $\bm{\textbf{IoU}\geq 0.7}$ \\
\midrule
100 & 0.4587 & 0.3629 & 0.2154 & 0.4510 & 0.3544 & 0.2073 & 0.4691 & 0.3646 & 0.2088 \\
200 & 0.4556 & 0.3631 & 0.2193 & 0.4510 & 0.3580 & 0.2142 & 0.4760 & 0.3759 & 0.2216 \\
300 & 0.4533 & 0.3552 & 0.2150 & 0.4494 & 0.3514 & 0.2113 & 0.4760 & 0.3709 & 0.2197 \\
400 & 0.4488 & 0.3471 & 0.2091 & 0.4452 & 0.3439 & 0.2062 & 0.4728 & 0.3642 & 0.2159 \\
500 & 0.4462 & 0.3425 & 0.2067 & 0.4428 & 0.3397 & 0.2036 & 0.4710 & 0.3606 & 0.2142 \\
\bottomrule
\end{tabular}
}
\label{tab:coarse_ablation_num_seg}
\end{table*}
%======================

%======================
\begin{table*}[t!]
\centering
\caption{\textbf{Ablation study on the usage of the pseudo training dataset for training the segment retrieval module.} Results from two CLIP models (using visual features from either a single frame CLIP$_{sf}$, or mean pooling from all frames CLIP$_{mp}$) serve as a reference, relying solely on the multi-modal alignment capability of the CLIP model without further training for segment retrieval. ``Top-20" and ``Top-40" refer to the number of video moments per query used for training from the pseudo training dataset. The threshold represents the segment's proportion of overlap with a moment annotation in the original TVR dataset. Note that the pseudo training dataset lacks manual annotations for the degree of relevance between a moment and the query. However, the moment annotations \ie start/end timestamps, are adopted from the original TVR dataset, hence are available in the TVR-Ranking dataset. Training with the top-40 samples and using only the segments whose overlap with moment annotation larger than 30\% yields the best results.}
\resizebox{1.0\linewidth}{!}{
\begin{tabular}{lccccccccccc}
\toprule
\multirow{2}{*}{\textbf{Model}} & \multirow{2}{*}{\textbf{Sample}} & \multirow{2}{*}{\textbf{Thresh.}} & \multicolumn{3}{c}{$\bm{\textbf{NDCG}@10}$} & \multicolumn{3}{c}{$\bm{\textbf{NDCG}@20}$} & \multicolumn{3}{c}{$\bm{\textbf{NDCG}@40}$} \\ 
\cmidrule(lr){4-6} \cmidrule(lr){7-9} \cmidrule(lr){10-12}
 & & & $\bm{\textbf{IoU}\geq 0.3}$ & $\bm{\textbf{IoU}\geq 0.5}$ & $\bm{\textbf{IoU}\geq 0.7}$ & $\bm{\textbf{IoU}\geq 0.3}$ & $\bm{\textbf{IoU}\geq 0.5}$ & $\bm{\textbf{IoU}\geq 0.7}$ & $\bm{\textbf{IoU}\geq 0.3}$ & $\bm{\textbf{IoU}\geq 0.5}$ & $\bm{\textbf{IoU}\geq 0.7}$ \\
\midrule
CLIP$_{sf}$ & - & - & 0.0271 & 0.0188 & 0.0083 & 0.0285 & 0.0198 & 0.0085 & 0.0317 & 0.0214 & 0.0089 \\
CLIP$_{mp}$ & - & - & 0.0364 & 0.0206 & 0.0110 & 0.0362 & 0.0203 & 0.0103 & 0.0386 & 0.0221 & 0.0106 \\
\midrule
\multirow{5}{*}{SP} & Top-20 & 0 & 0.3902 & 0.2649 & 0.1291 & 0.3854 & 0.2601 & 0.1250 & 0.4068 & 0.2714 & 0.1274 \\
 & Top-20 & 0.3 & 0.4120 & 0.3196 & 0.1877 & 0.4043 & 0.3136 & 0.1824 & 0.4267 & 0.3289 & 0.1879 \\
 & Top-40 & 0 & 0.4327 & 0.3033 & 0.1636 & 0.4322 & 0.2970 & 0.1624 & 0.4555 & 0.3117 & 0.1668 \\
 & Top-40 & 0.3 & 0.4556 & 0.3631 & 0.2193 & 0.4510 & 0.3580 & 0.2142 & 0.4760 & 0.3759 & 0.2216 \\
 & Top-40 & 0.5 & 0.4313 & 0.3364 & 0.2003 & 0.4292 & 0.3374 & 0.1937 & 0.4545 & 0.3469 & 0.1982 \\
\bottomrule
\end{tabular}
}
\label{tab:coarse_ablation_training_samples}
\end{table*}
%======================

%======================
\section{Ablation Study on Segment Retrieval}
%======================
We conduct two ablation studies related to the segment retrieval module. The first evaluates the impact of the number of retrieved segments on the final results. The second investigates the best approach to utilize the pseudo-training set in TVR-Ranking to develop the most effective segment retrieval.

%=============================
\subsection{Number of Segments to Retrieve}
%=============================
In our framework, the segments retrieved in the first step serve as inputs to the subsequent modules, \ie proposal generation, and refinement and re-ranking. If too few segments are retrieved, they limit the number of moments that can be located in the next phases. Conversely, retrieving too many segments introduces unnecessary noise to the subsequent modules and impacts inference efficiency.

To examine the impact, we report results from SP with a flat index setting, varying the number of retrieved segments from 100 to 500. Results in \cref{tab:coarse_ablation_num_seg} indicate that performance peaks at 200 segments across most metrics. Upon further analysis, we find that retrieving more than 200 segments leads to additional segments \textit{around the target moments} being retrieved. These additional segments do not lead to the retrieval of new target moments. Instead, they decrease the IoU with the target moments in the generated proposals, leading to additional computing overhead. In other words, if a target moment is missed in the top 200 retrieved segments, the chance of retrieving it is not significantly increased by further increasing the number of retrieved segments. Therefore, we set the retrieval number to 200 segments by default in all experiments to ensure optimal performance and efficiency.

%======================
\subsection{Pseudo-training Set for Segment Retrieval}
%======================

In the segment retrieval stage, we train a pair of feature projectors to align video segments and text queries in a shared feature space.

There are two options here: The first is to utilize multi-modality base models, such as CLIP~\cite{radford2021learning}, which are pre-trained with multi-modal alignment. In this case, no additional training is required. The second option is to use the pseudo-training set provided in the TVR-Ranking dataset to develop a segment retrieval module tailored to the domain-specific types of videos and queries in TVR-Ranking.

Regarding the choice of using multi-modal base models without additional training, we experimented with two configurations of CLIP. The first samples the middle frame of each segment and uses the frame embedding as the segment embedding. The second generates image embeddings for all frames and obtains the segment embedding through mean pooling. We denote these versions as $\text{CLIP}_{sf}$ and $\text{CLIP}_{mp}$, respectively.

In TVR-Ranking, the pseudo-training set lacks \textit{manual} annotations for the \textit{degree of relevance} between a moment and the query. However, the moment annotations (\ie start/end timestamps) are adopted from the original TVR dataset and are thus available in TVR-Ranking. Based on the semantic similarity between the query and the re-written moment descriptions derived from TVR, TVR-Ranking provides up to 40 relevant moments for each query in its pseudo-training set.

We evaluated several configurations for utilizing the pseudo-training set. First, we tested training with either the top 20 or top 40 relevant moments. Then, when a video is divided into segments, the annotated moments from TVR may be split across multiple segments, some of which may have very little overlap with the moment. We apply a threshold to exclude segments where the proportion of overlap is below a specified value (\eg 0.3 or 0.5).

From the  results presented in \cref{tab:coarse_ablation_training_samples}, we make the following observations. First, although the CLIP model provides strong visual-text alignment, its performance is suboptimal without appropriate training. This highlights the importance of learning temporal modeling and adapting to the downstream dataset to create an effective segment-query aligner. Additionally, training on a larger set of pseudo-positive samples does lead to higher performance. Finally, using a threshold of 0.3 yields the best results for trained projectors, outperforming thresholds of 0 (\ie no filtering of segment) and 0.5 (\ie filtering too many segments). Based on these findings, we choose to train the model by using the top-40 relevant moments in pseudo-training  with a filter threshold of 0.3.

\end{document}